\documentclass[english,aps,manuscript,prb,a4paper,twocolumn,reprint,noshowpacs,superscriptaddress,floatfix]{revtex4-1}
\usepackage[T1]{fontenc}
\usepackage[utf8]{inputenc}
\usepackage[a4paper]{geometry}
\usepackage{color}

\usepackage{array}
\usepackage{mathrsfs}
\usepackage{multirow}
\usepackage{amsmath,amssymb}
\usepackage{graphicx}
\usepackage{esint}

\makeatletter

\providecommand{\tabularnewline}{\\}

\@ifundefined{textcolor}{}
{%
 \definecolor{BLACK}{gray}{0}
 \definecolor{WHITE}{gray}{1}
 \definecolor{RED}{rgb}{1,0,0}
 \definecolor{GREEN}{rgb}{0,1,0}
 \definecolor{BLUE}{rgb}{0,0,1}
 \definecolor{CYAN}{cmyk}{1,0,0,0}
 \definecolor{MAGENTA}{cmyk}{0,1,0,0}
 \definecolor{YELLOW}{cmyk}{0,0,1,0}
}

\makeatother

\usepackage{babel}
\begin{document}

\title{Spin wave approach to the two-magnon Raman scattering in an $J_{1x}-J_{1y}-J_{2}-J_{c}$
antiferromagnetic Heisenberg model}

\author{Changle Liu}
\affiliation{Department of Physics and Beijing Key Laboratory of Opto-electronic Functional Materials \& Micro-nano Devices, Renmin University of China, Beijing 100872, China
}
\author{Rong Yu}
\affiliation{Department of Physics and Beijing Key Laboratory of Opto-electronic Functional Materials \& Micro-nano Devices, Renmin University of China, Beijing 100872, China
}
\affiliation{Department of Physics and Astronomy, Collaborative Innovation
Center of Advanced Microstructures, Shanghai Jiaotong University, Shanghai
200240, China}

\date{\today}
\begin{abstract}
We study the two-magnon non-resonant Raman scattering in the $(\pi,\pi)$
and $(\pi,0)$ ordered antiferromagnetic phases of a $J_{1x}-J_{1y}-J_{2}-J_{c}$ Heisenberg model on the tetragonal
lattice within the framework of the spin-wave theory. We discuss the effects of various tuning factors to the two-magnon Raman spectra. We find that both the magnetic frustration $J_2/J_1$ and the interlayer exchange coupling $J_c$ may  significantly affect the spectra in both the $B_{1g}$ and $A_{1g}^\prime$ channels in the $(\pi,\pi)$ N\'{e}el ordered phase. Moreover, we find a splitting of the two-magnon peak in the $(\pi,0)$ antiferromagnetic phase. We further discuss the implications of our results to the BaMnBi$_2$ and iron pnictide systems.
\end{abstract}
\maketitle

\section{Introduction}

In the recent years, the discovery of iron-based superconductors have
triggered tremendous research in this new class of high $T_{c}$ superconductors. Similar to cuprates, iron-based superconductors have a layered structure,
and their parent compounds have long-range antiferromagnetic order.
Superconductivity emerges when the magnetic order is destroyed by doping.
It is now believed that magnetism is crucial for superconductivity in these materials.
Typically the parent compounds of iron pnictides have a $(\pi,0)$
collinear antiferromagnetic order. The magnetic properties of these
materials can be well captured by 
the strong coupling approaches involving interactions
of Fe local spins, described by effective $J_1-J_2$ like models. These extended antiferromagnetic Heisenberg models are widely used to explain the magnetic properties of parent iron pnictides\cite{zhao2009spin}.

There are many other materials which invoke the local moment models.
Recently, a class of novel manganese based materials $A\mathrm{MnBi_{2}}$ ($A=\mathrm{Sr,Ca}$) have
attracted considerable research interests for their coexistence of
itinerant Dirac electrons and long-range magnetic order associated with local moments. These materials share similar
structrual and electronic properties to iron pnictides. Insulating$\mathrm{MnBi}$
layer with N\'{e}el-type antiferromagnetic order on each Mn site\cite{guo2014coupling}
and $A\mathrm{Bi}$ layer accomodating highly anisotropic Dirac carriers\cite{feng2014strong,guo2014coupling,he2012giant,jia2014observation,lee2013anisotropic,park2011anisotropic,wang2011layered,wang2011quantum,wang2012two}
are alternatively stacked. These materials have provided an opportunity
to explore the interplay of magnetism and Dirac itinerant carriers.

In studying the magnetic properties of these systems, Raman scattering is a powerful spectroscopic technique. It probes two-magnon correlations in which
short-wavelength excitations dominates. The standard magnetic Raman
scattering theory is based on the Fleury-Loudon coupling between the light
and the spin system\cite{fleury1968scattering}. Such a theory can
be derived in the large-U Hubbard model at half-filling in the non-resonant
regime\cite{PhysRevLett.65.1068}. Near resonance where the
incoming photon frequency is close to the band gap value, FL theory
fails as the charge transfer process becomes dominant\cite{chubukov1995resonant}.

Even in the non-resonant
regime, theoretical understanding of Raman scattering in spin systems is still very limited. Early works on the simple 2D antiferromagnets have revealed that magnon-magnon (m-m) interactions have significant influence on the shape of Raman spectra as
multiple scattering of magnon pairs excited by photons is non-negligible
in the Raman process\cite{davies1971spin}. Magnetic Raman scattering
in 2D simple antiferromagnets has been further studied using various
approaches: spin wave \& Green's function theory\cite{canali1992theory,sandvik1998numerical,suzuki1992application},
Exact Diagonalization (ED), Quantum Monte Carlo method\cite{sandvik1998numerical}, etc.. Ref. \cite{canali1992theory}
claimed that four-magnon intensity is too small compared with two-magnon
ones. \cite{suzuki1992application} calculated both two-magnon and
four magnon Raman spectra for a 2D frustrated $J_{1}-J_{2}$ systems
with N\'{e}el order using the modified spin wave (MSW) theory. However, their
calculations were still at the mean-field level, which ignored higher
order scattering processes of magnon pairs. Ref. \cite{katanin2003theoretical}
calculated 2D systems with ring exchange interactions. Ref. \cite{chen2011theory}
launched ED calculations for specific iron-based materials in $(\pi,0)$
collinear and $(\pi/2,\pi/2)$ diagonal double stripe order. However,
their calculations were restricted to small spin ($S\leq1$), small
cluster sizes ($N_{c}\leq36$) and very limited system parameters.

Up to now, a convincing and detailed work that is applicable to systems
with frustration, exchange anisotropy, and finite interlayer exchange couplings is still absent. In this article we present a systematic study of the two-magnon non-resonant
Raman scattering in $(\pi,\pi)$ and $(\pi,0)$ ordered antiferromagnets
with square/tetragonal \textit{lattice} geometry (notice that the
magnetic symmetry can be lower than lattice symmetry) within the framework
of spin-wave theory. We will discuss its general feature,
and implications of spin magnitude, frustration, anisotropy and interlayer
coupling to Raman spectra. The article is organized as follows. In
Sec. II we introduce our calculation method. In Sec. III and IV we
present our results in $(\pi,\pi)$ and $(\pi,0)$ ordered system
respectively. Finally in Sec. IV we present our discussions and concluding
remarks.

\section{General Formalism}

We study 
a spin-$S$ $J_{1x}-J_{1y}-J_{2}-J_{c}$ Heisenberg model on a tetragonal lattice.
The
Hamiltonian reads

\begin{eqnarray}
H & = & \frac{J_{1x}}{2}\sum_{i,\delta_{x}}\mathbf{S}_{i}\cdot S_{i+\delta_{x}}+\frac{J_{1y}}{2}\sum_{i,\delta_{y}}\mathbf{S}_{i}\cdot\mathbf{S}_{i+\delta_{y}}\nonumber \\
 & + & \frac{J_{2}}{2}\sum_{i,\delta_{x},\delta_{y}}\mathbf{S}_{i}\cdot\mathbf{S}_{i+\delta_{x}+\delta_{y}}+\frac{J_{c}}{2}\sum_{i,\delta_{z}}\mathbf{S}_{i}\cdot\mathbf{S}_{i+\delta_{z}}\label{eq:ham}
\end{eqnarray}
where $\mathbf{S}_i$ refers to a spin at lattice site $i$,
and $\delta_{x}=\pm a\hat{x}$,$\delta_{y}=\pm a\hat{y}$,
$\delta_{z}=\pm c\hat{z}$ are nearest neighbor vectors along x, y,
z directions, respectively. $J_{1x}$, $J_{1y}$, $J_2$, and $J_c$ 
are, respectively, the exchange couplings
between row and column nearest neighbors, next nearest neighbors, and
nearest interlayer neighbors. In this paper, we are interested in the Raman scattering with an AF ground state. Without losing generality, we take $J_{1x}>0$ in this model.

\begin{figure}
\includegraphics[scale=0.75]{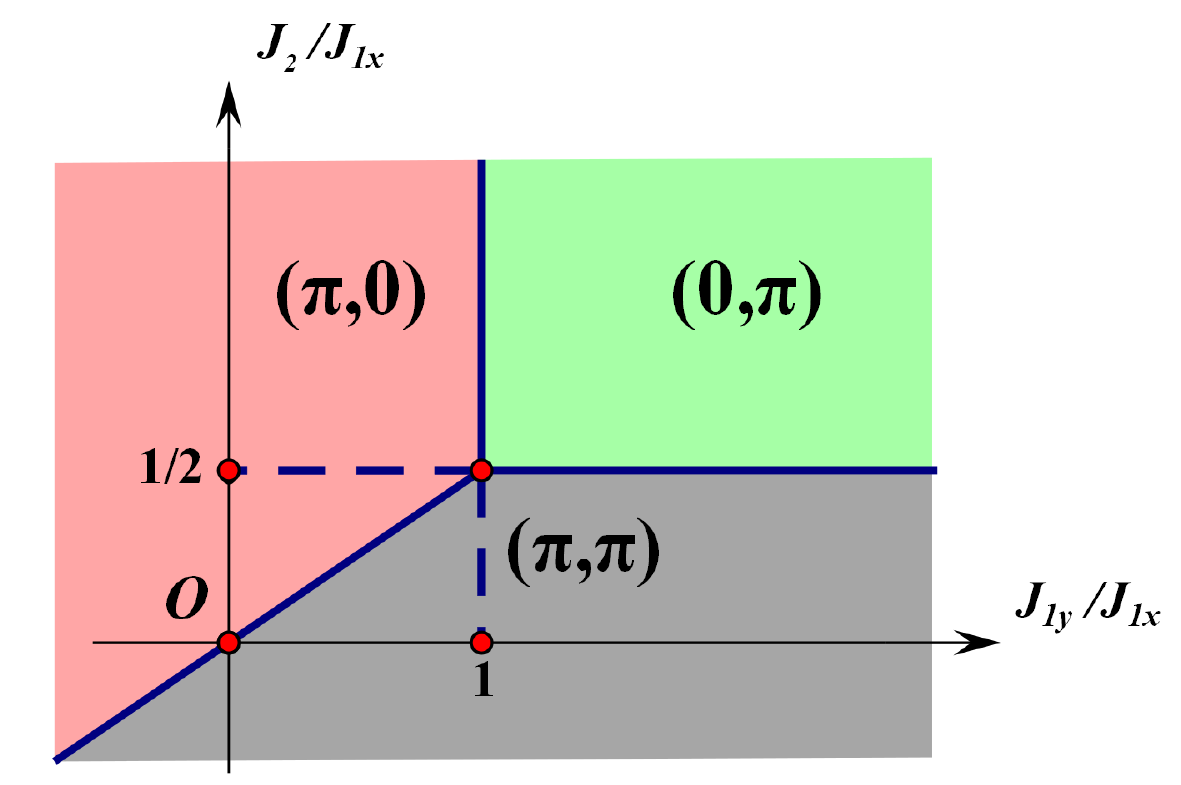}

\protect\caption{(Color online). Ground state phase diagram of the classical $J_{1x}-J_{1y}-J_{2}$ model in the regime $J_{1x}>0$, $J_{1y}/J_{1x}\geqslant-1$ and $J_{1y}+2J_{2}\geqslant0$, where the ground state stabilizes in-plane AFM orders with wave vectors $(\pi,0)$, $(0,\pi)$, and $(\pi,\pi)$, respectively. The solid lines mark the phase boundaries. 
}

\label{fig:phase}
\end{figure}

The ground state phase diagram of the classical $J_{1x}-J_{1y}-J_2$ model is illustrated in
Fig. \ref{fig:phase}. In the regime shown, three in-plane AFM ground states can be stabilized.  In each of the three ordered states, the quantum fluctuations of the corresponding quantum spin model are taken into account by the spin-wave approach via a standard $1/S$ expansion, which 
is expected to be a good approximation when the system is not in the vicinity of the classical phase boundaries.

\subsection{$1/S$ Expansion}

By introducing Holstein-Primakoff (H-P) transformation on the bipartite tetragonal lattice, we express
spins in the A (spin up) and B (spin down) sublattices in terms of bosonic
operators $a_{l}$ and $b_{m}$

\begin{equation}
\begin{split}S_{l}^{z} & =S-a_{l}^{\dagger}a_{l}\\
S_{l}^{+} & =\sqrt{2S}f_{l}(S)a_{l}\\
S_{m}^{z} & =-S+b_{m}^{\dagger}b_{m}\\
S_{m}^{+} & =\sqrt{2S}b_{m}^{\dagger}f_{m}(S),
\end{split}
\end{equation}
where
\begin{equation}\label{Eq:constraint}
f_{l/m}(S)=\sqrt{1-\frac{n_{l/m}}{2S}},
\end{equation}
$n_{l/m}=a^\dagger_{l/m} a_{l/m}$, and $l\in A$ and $m\in B$. For Eq.~\eqref{Eq:constraint} to be valid, the bosons must be
restricted in the $n_{l/m}\leq2S$
physical space.

Then we perform an $1/S$ expansion for $f_{l/m}(S)$ up to the $1/S$ order:

\begin{equation}
f_{l/m}(S)=1-\frac{n_{l/m}}{4S}+...,
\end{equation}
and perform a Fourier transformation for the bosonic operators

\begin{equation}
\begin{split}a_{l}=\sqrt{\frac{2}{N}}\sum_{\mathbf{k}}a_{\mathbf{k}}e^{i\mathbf{\mathbf{k}\cdot\mathbf{R_{l}}}}, & \: b_{m}=\sqrt{\frac{2}{N}}\sum_{\mathbf{k}}b_{-\mathbf{k}}e^{-i\mathbf{k}\cdot\mathbf{R_{m}}}\end{split}
\end{equation}
where $\mathbf{k}$ is defined in the first Brillouin zone (FBZ) of the momemtum space.

The Hamiltonian is also expanded in powers of $1/S$ as
\begin{equation}
H=E_{0}+H_{0}+H_{1}+ O(1/S^2). 
\label{eq:ham_in_S}
\end{equation}

Here, $E_{0}$ corresponds to the classical
energy of the system. $H_{0}$ corresponds to the quadratic linear spin
wave (LSW) terms, which takes the form

\begin{equation}
H_{0}=\sum_{\mathbf{k}}P_{\mathbf{k}}(a_{\mathbf{k}}^{\dagger}a_{\mathbf{k}} + b_{-\mathbf{k}}^{\dagger}b_{-\mathbf{k}}) + Q_{\mathbf{k}}(a_{\mathbf{k}}^{\dagger}b_{-\mathbf{k}}^{\dagger} + a_{\mathbf{k}}b_{-\mathbf{k}}),
\end{equation}
where the coefficients $P_{\mathbf{k}}$ and $Q_{\mathbf{k}}$ are defined in the Appendix.
Next we perform Bogoliubov transformation
\begin{equation}
\begin{aligned}a_{\mathbf{k}}^{\dagger}=l_{\mathbf{k}}\alpha_{\mathbf{k}}^{\dagger}+m_{\mathbf{k}}\beta_{-\mathbf{k}},\; & b_{-\mathbf{k}}=m_{\mathbf{k}}\alpha_{\mathbf{k}}^{\dagger}+l_{\mathbf{k}}\beta_{-\mathbf{k}}\end{aligned}
\end{equation}
where $l_{\mathbf{k}}=\sqrt{\frac{1+\epsilon_{\mathbf{k}}}{2\epsilon_{\mathbf{k}}}}$,
$m_{\mathbf{k}}=-x_{\mathbf{k}}l_{\mathbf{k}}=-\mathrm{sgn}\gamma_{\mathbf{k}}\sqrt{\frac{1-\epsilon_{\mathbf{k}}}{2\epsilon_{\mathbf{k}}}}$,
$\epsilon_{\mathbf{k}}=\sqrt{1-\gamma_{\mathbf{k}}^{2}}$, $\gamma_{\mathbf{k}}=Q_{\mathbf{k}}/P_{\mathbf{k}}$.
$H_{0}$ is then diagonalized as
\begin{equation}
H_{0}=\sum_{\mathbf{k}}\omega_{\mathbf{k}}(\alpha_{\mathbf{k}}^{\dagger}\alpha_{\mathbf{k}}+\beta_{-\mathbf{k}}^{\dagger}\beta_{-\mathbf{k}}+1)-P_{\mathbf{k}},
\end{equation}
where $\omega_{\mathbf{k}}=P_{\mathbf{k}}\epsilon_{\mathbf{k}}$.

$H_{1}$ corresponds to the $1/S$ order correction to the LSW results. It
is written in Bogoliubov magnons as
\begin{equation}\label{eq:H1}
H_{1}=const.+H_{0}^\prime+H_{1}^\prime+...
\end{equation}
where
\begin{equation}
H_{0}^\prime=\sum_{\mathbf{k}}A_{\mathbf{k}}(\alpha_{\mathbf{k}}^{\dagger}\alpha_{\mathbf{k}}+\beta_{\mathbf{k}}^{\dagger}\beta_{\mathbf{k}})+B_{\mathbf{k}}(\alpha_{\mathbf{k}}^{\dagger}\beta_{-\mathbf{k}}^{\dagger}+\alpha_{\mathbf{k}}\beta_{-\mathbf{k}}),
\end{equation}
is known as the Oguchi correction 
arising from transforming the
bosonic operators into normal products. The Oguchi terms give the $1/S$ order
correction to the magnon dispersion $\tilde{\omega}_{\mathbf{k}}=\omega_{\mathbf{k}}+A_{\mathbf{k}}$.

\begin{eqnarray}
H_{1}^\prime & = & \frac{2}{N}\sum_{1234}\delta_{\mathbf{G}}(1+2-3-4)l_{1}l_{2}l_{3}l_{4}[B_{1234}^{(1)}\alpha_{1}^{\dagger}\alpha_{2}^{\dagger}\alpha_{3}^{\dagger}\alpha_{4}^{\dagger}+\nonumber \\
 &  & B_{1234}^{(2)}\beta_{-3}^{\dagger}\beta_{-4}^{\dagger}\beta_{-1}\beta_{-2}+B_{1234}^{(3)}\alpha_{1}^{\dagger}\beta_{-4}^{\dagger}\beta_{-2}\alpha_{3}+\nonumber \\
 &  & (B_{1234}^{(4)}\alpha_{1}^{\dagger}\beta_{-2}\alpha_{3}\alpha_{4}+B_{1234}^{(5)}\beta_{-4}^{\dagger}\beta_{-1}\beta_{-2}\alpha_{3}+\nonumber \\
 &  & B_{1234}^{(6)}\alpha_{1}^{\dagger}\alpha_{2}^{\dagger}\beta_{-3}^{\dagger}\beta_{-4}^{\dagger}+h.c.)]
\end{eqnarray}
where 1, 2, 3, 4 are abbreviations of the wave vectors $\mathbf{k}_{1}$, $\mathbf{k}_{2}$,
$\mathbf{k}_{3}$, $\mathbf{k}_{4}$, which are also defined in FBZ,
$\delta_{\mathbf{G}}(1+2-3-4)$ represents the conservation of momenta
within a reciprocal lattice vector $\mathbf{G}$.$H_0^\prime$ consists of two-magnon scattering terms of the magnon-magnon (m-m) interaction depending on the coefficients $P_{\mathbf{k}}$, $Q_{\mathbf{k}}$, $A_{\mathbf{k}}$, $B_{\mathbf{k}}$
and the vertex factor $B_{1234}^{(3)}$, whose explicit forms are given in Appendix A.

\subsection{Two-magnon Raman operator}

In the standard magnetic Raman FL theory, the second order Raman scattering
operator is given by\cite{fleury1968scattering}

\begin{equation}
\hat{O}=\frac{\lambda}{2}\sum_{ij}J_{ij}(\hat{\mathbf{e}}_{in}\cdot\hat{\mathbf{d}}_{ij})(\hat{\mathbf{e}}_{out}\cdot\hat{\mathbf{d}}_{ij})\mathbf{S}_{i}\cdot\mathbf{S}_{j}\label{eq: FL}
\end{equation}
where $\mathbf{\hat{e}}_{in}$ and $\mathbf{\hat{e}}_{out}$ are unit
polarization vectors of the incoming and scattered lights. $\mathbf{\hat{d}}_{ij}$
is the vector connecting site $i$ and site $j$. $\lambda$ is the
coupling constant, which is scaled to $\sqrt{\frac{2}{N}}$, as its
magnitude is not important for our results.

We will consider the following light polarization geometries which are widely accepted in experimental set up:
$\mathbf{\hat{e}}_{in}=\frac{1}{\sqrt{2}}(\hat{x}+\hat{y})$, $\mathbf{\hat{e}}_{out}=\frac{1}{\sqrt{2}}(\hat{x}-\hat{y})$
for $x^\prime y^\prime $ polarization,
and
$\mathbf{\hat{e}}_{in}=\frac{1}{\sqrt{2}}(\hat{x}+\hat{y})$, $\mathbf{\hat{e}}_{out}=\frac{1}{\sqrt{2}}(\hat{x}+\hat{y})$
for $x^\prime x^\prime $ polarization. Here $x^\prime $ and $y^\prime $ refer to the rotated axes after a 45$^\circ$ rotation about the $z$ axis in the $xy$ plane.

Note that if the system has $D_{4h}$ symmetry, $x'y'$ polarization corresponds
to $B_{1g}\bigoplus A_{2g}$ symmetry group representations, and is
usually called the $B_{1g}$ channel 
because the $A_{2g}$ component
is zero in the second-order Raman scattering. The $x'x'$ polarization
corresponds to $A_{1g}\bigoplus B_{2g}$ representations and is 
denoted as the $A_{1g}'$ channel. In $D_{4h}$ symmetric system
these two channels are well separated from a symmetry perspective. If the $D_{4h}$ symmetry is
broken, they may not refer to the different irreducible representations of the symmetry group, and as a consequence, the scattering signals from these two channels can mix. 

At the LSW level, the two-magnon part of the Raman operator is given by
\begin{equation}
\hat{O}=\sum_{\mathbf{k}}M_{\mathbf{k}}(\alpha_{\mathbf{k}}^{\dagger}\beta_{-\mathbf{k}}^{\dagger}+\alpha_{\mathbf{k}}\beta_{-\mathbf{k}})+...
\end{equation}
where explicit forms of $M_{\mathbf{k}}$ in $B_{1g}$ and $A_{1g}'$
channels are given in Appendix A.

\subsection{Raman Scattering Cross Section}

The Raman scattering cross section at zero temperature is given by $R(\omega)=-\frac{1}{\pi}\mathbf{\mathrm{Im}}[I(\omega)]$,
where $I(\omega)$ is the correlation function $I(\omega)=-i\int\mathrm{d}t\, e^{i\omega t}\langle\mathscr{T}_{t}\,\hat{O}^{\dagger}(t)\hat{\, O}(0)\rangle_{0}$.
Here $\langle...\rangle_{0}$ represents quantum mechanical average
over the ground state, and $\mathscr{T}_{t}$ is the time ordering
operator. The two-magnon contribution to $I(\omega)$ can be written
as $I(\omega)=\sum_{\mathbf{k,k'}}M_{\mathbf{k}}\Pi_{\mathbf{kk'}}(\omega)M_{\mathbf{k'}}$.
Here we define two-magnon Green's function
\begin{equation}
\Pi_{\mathbf{kk'}}(\omega)=-i\int\mathrm{d}t\, e^{i\omega t}\langle\mathscr{T_{t}}\,\alpha_{k}(t)\beta_{-k}(t)\alpha_{k'}^{\dagger}(0)\beta_{-k'}^{\dagger}(0)\rangle_{0}\label{eq:gf}
\end{equation}

To calculate $I(\omega)$ we have adopted the following two  simplifications\cite{canali1992theory,luo2014spectrum}:

1) We expand the one-magnon propagator
up to the $1/S$ order. 
Given that there is no correction to the propagator at the $1/S$
order, 
they 
are identical
to the unperturbed ones:
\begin{equation}
\begin{split}G_{\alpha\alpha}(\mathbf{k},\omega) & =G_{\beta\beta}(\mathbf{k},\omega)=\frac{1}{\omega+i\,0^{+}-\tilde{\omega}_{\mathbf{k}}}\\
G_{\alpha\beta}(\mathbf{k},\omega) & =G_{\beta\alpha}(\mathbf{k},\omega)=0
\end{split}
\end{equation}
By applying Wick's theorem, the unperturbed two-magnon propagator
can be expanded 
in terms of one-magnon ones as
\begin{eqnarray}
\Pi_{\mathbf{kk'}}^{(0)}(\omega) & = & \delta_{\mathbf{kk'}}i\int\frac{\mathrm{d}\omega_{1}}{2\pi}G_{\alpha\alpha}(\mathbf{k},\omega+\omega_{1})G_{\beta\beta}(-\mathbf{k},-\omega_{1})\nonumber \\
 & = & \delta_{\mathbf{kk'}}\frac{1}{\omega+i\,0^{+}-2\tilde{\omega}_{\mathbf{k}}}
\end{eqnarray}

\begin{figure}
\includegraphics[scale=0.5]{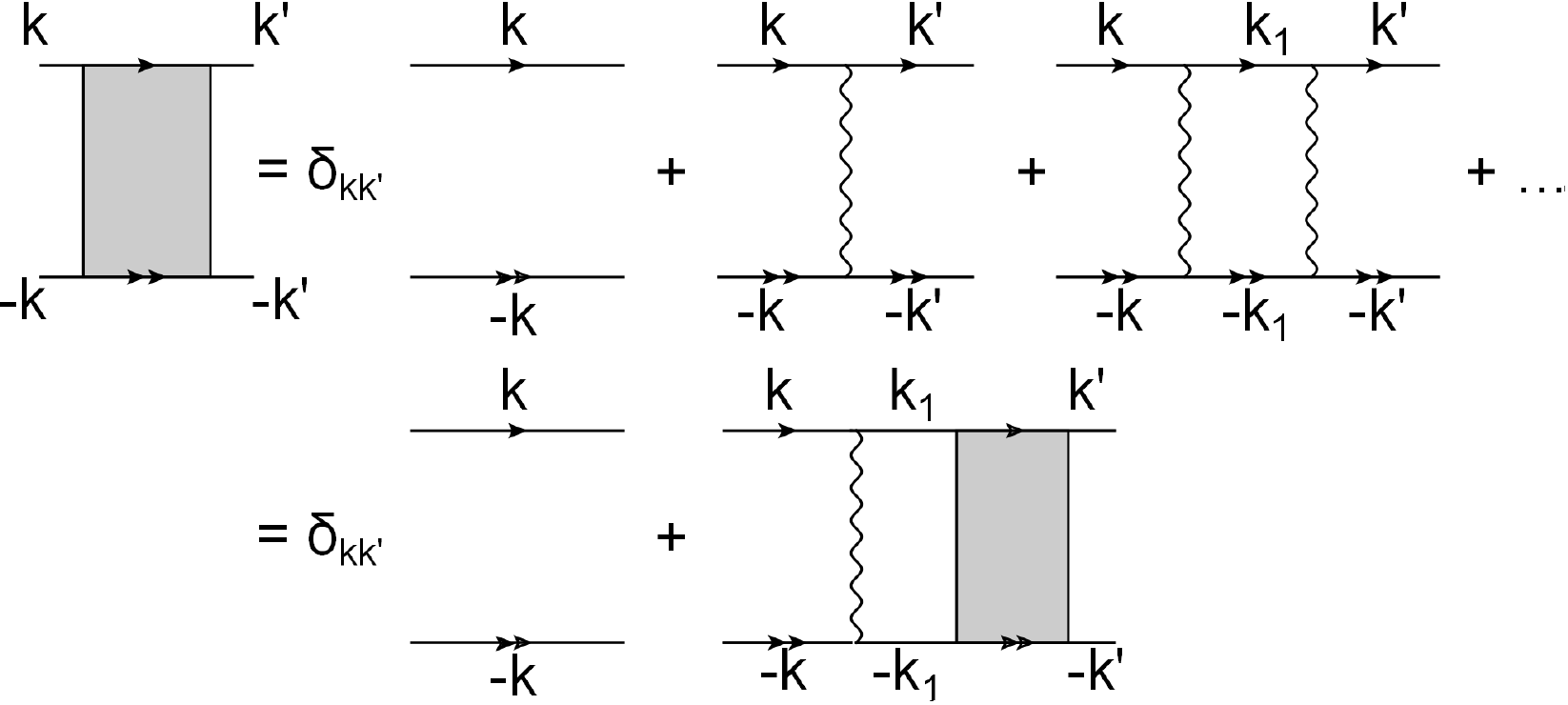}

\protect\caption{(Color online). Ladder diagrams for two-magnon propagator (\ref{eq:gf}).
A line with single arrow represents $\alpha$ magnon propagator $G_{\alpha\alpha}(\mathbf{k},\omega)$.
A line with double arrow represents $\beta$ magnon propagator $G_{\beta\beta}(\mathbf{k},\omega)$. }

\label{fig:ladder}
\end{figure}

2) The m-m interaction is taken into account 
within the framework of the ladder approximation for the two-magnon Green's function (\ref{eq:gf}). In this approximation, the core vertex is expanded up to the $1/S$ order, 
and only vertex terms equivalent to $\alpha^{\dagger}\beta^{\dagger}\beta\alpha$ are kept,
as illustrated in Fig. \ref{fig:ladder}. As the total
momentum of incoming and outgoing $\alpha$ and $\beta$ magnons are
fixed to 0, the vertex is the function of the incoming and outgoing
$\alpha$ magnon momentum $\mathbf{k}$ and $\mathbf{k'}$, i.e. $\mathscr{V}_{\mathbf{kk'}}=\frac{2}{N}B_{\mathbf{k'kkk'}}^{(3)}$.

Therefore we have
\begin{equation}
\Pi_{\mathbf{kk'}}(\omega)=i\int\frac{d\omega'}{2\pi}G_{\alpha\alpha}(\mathbf{k},\omega+\omega')G_{\beta\beta}(-\mathbf{k},-\omega')\Gamma_{\mathbf{kk}'}(\omega,\omega')\label{eq:BS1}
\end{equation}
where the vertex function $\Gamma_{\mathbf{kk'}}(\omega,\omega')$
satisfies Bathe-Salpeter equation
\begin{eqnarray}
\Gamma_{\mathbf{kk'}}(\omega,\omega') & = & \delta_{\mathbf{kk'}}+i\sum_{\mathbf{k_{1}}}\int\frac{d\omega_{1}}{2\pi}\mathscr{V}_{\mathbf{kk_{1}}}G_{\alpha\alpha}(\mathbf{k}_{1},\omega+\omega_{1})\nonumber \\
 & \times & G_{\beta\beta}(-\mathbf{k}_{1},-\omega_{1})\Gamma_{\mathbf{k_{1}k'}}(\omega,\omega_{1})\label{eq:BS2}
\end{eqnarray}
Since $\mathscr{V}_{\mathbf{kk'}}$ is independent of frequencies.
From the equation (\ref{eq:BS2}), $\Gamma_{\mathbf{kk'}}(\omega,\omega')$
is independent of $\omega'$, i.e. $\Gamma_{\mathbf{kk'}}(\omega,\omega')=\Gamma_{\mathbf{kk'}}(\omega)$

In Eqs.~\eqref{eq:BS1} and~\eqref{eq:BS2}, integration over frequencies are
decoupled
\begin{eqnarray}
\Pi_{\mathbf{kk'}}(\omega) & = & \Pi_{\mathbf{kk}}^{(0)}(\omega)\Gamma_{\mathbf{kk'}}(\omega)\nonumber \\
\Gamma_{\mathbf{kk'}}(\omega) & = & \delta_{\mathbf{kk'}}+\sum_{\mathbf{k_{1}}}\mathscr{V}_{\mathbf{kk_{1}}}\Pi_{\mathbf{k_{1}k_{1}}}^{(0)}(\omega)\Gamma_{\mathbf{k_{1}k'}}(\omega)
\end{eqnarray}
Eliminating $\Gamma$ vertex, we get two-magnon Dyson's equation:
\begin{equation}
\hat{\Pi}=\hat{\Pi}^{(0)}+\hat{\Pi}^{(0)}\hat{\mathscr{V\:}}\hat{\Pi}=\hat{\Pi}^{(0)}\sum_{n=0}^{+\infty}(\hat{\mathscr{V\:}}\hat{\Pi}^{(0)})^{n}\label{eq: dyson}
\end{equation}
Directly solving such an equation rigorously $\hat{\Pi}=[\hat{1}-\hat{\Pi}^{(0)}\hat{\mathscr{V\:}}]^{-1}\hat{\Pi}^{(0)}$
would require the inverse of the matrix with a $N/2\times N/2$ dimension, which
is obviously computationally expensive. So we use the following alternative approach:
The vertex function can be expressed as a separable form
\begin{equation}
\mathscr{V}_{\mathbf{kk'}}=\sum_{m,n=1}^{Nc}v_{m\mathbf{k}}\Gamma_{mn}v_{n\mathbf{k}}\label{eq:saperateV}
\end{equation}
where explicit forms of $\hat{\Gamma}$ and $v_{n\mathbf{k}}$ are
given in Appendix A. We have
\begin{eqnarray}
\hat{\Pi} & = & \hat{\Pi}^{(0)}\sum_{n=0}^{+\infty}(\hat{v}^{T}\hat{\Gamma}\hat{v}\hat{\Pi}^{(0)})^{n}\nonumber \\
 & = & \hat{\Pi}^{(0)}+(\hat{v}\hat{\Pi}^{(0)})^{T}\cdot\hat{\Gamma}\sum_{n=0}^{+\infty}[(\hat{v}\hat{\Pi}^{(0)}\hat{v}^{T})\hat{\Gamma}]^{n}\cdot(\hat{v}\hat{\Pi}^{(0)})\nonumber \\
 & = & \hat{\Pi}^{(0)}+\hat{\Pi}^{L}
\end{eqnarray}
where
\begin{eqnarray}
\hat{\Pi}^{L} & = & (\hat{v}\hat{\Pi}^{(0)})^{T}\cdot\hat{\Gamma}[\mathbf{1}-(\hat{v}\hat{\Pi}^{(0)}\hat{v}^{T})\hat{\Gamma}]^{-1}\cdot(\hat{v}\hat{\Pi}^{(0)})
\end{eqnarray}
is the ladder correction to the two-magnon propagators.

Thus we have 
obtained an approach of exactly solving the Dyson's equation
(\ref{eq: dyson}) with the price of inverting matrix with mere dimension
of $N_{c}\times N_{c}$. Finally, the correlation function $I(\omega)$
can be obtained by
\begin{eqnarray}
I(\omega) & = & I^{(0)}(\omega)+I^{L}(\omega)
\end{eqnarray}
where
\begin{eqnarray}
I^{(0)}(\omega) & = & \hat{M^{T}}\hat{\Pi}^{(0)}\hat{M}
\end{eqnarray}
and
\begin{eqnarray}
I^{L}(\omega) & = & \hat{M^{T}}\hat{\Pi}^{L}\hat{M}=(\hat{v}\hat{\Pi}^{(0)}\hat{M})^{T}\nonumber \\
 & \cdot & \hat{\Gamma}[\hat{\mathbf{1}}-(\hat{v}\hat{\Pi}^{(0)}\hat{v}^{T})\hat{\Gamma}]^{-1}\cdot(\hat{v}\hat{\Pi}^{(0)}\hat{M})
\end{eqnarray}
are non-interacting and ladder corrections to the 
total scattering cross section, respectively.

\section{Results For the $(\pi,\pi)$ Néel order}
We first consider the results of Raman scattering when the ground state has an AFM N\'{e}el order at wave vector $(\pi,\pi)$. We discuss several factors that may affect the Raman spectrum.

\subsection{Role of $1/S$}

\begin{figure}
\includegraphics[clip,scale=0.5]{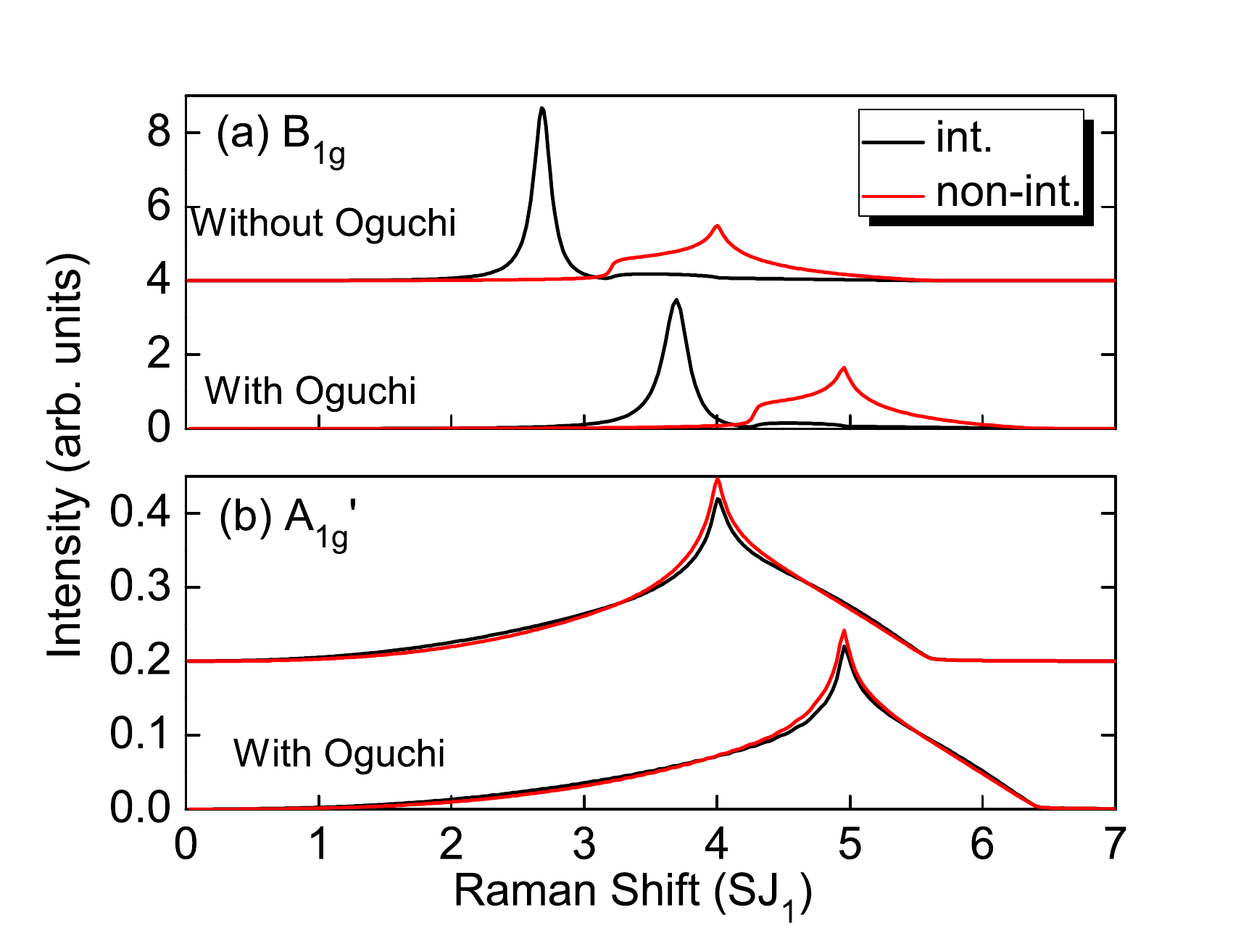}
\includegraphics[clip,scale=0.5]{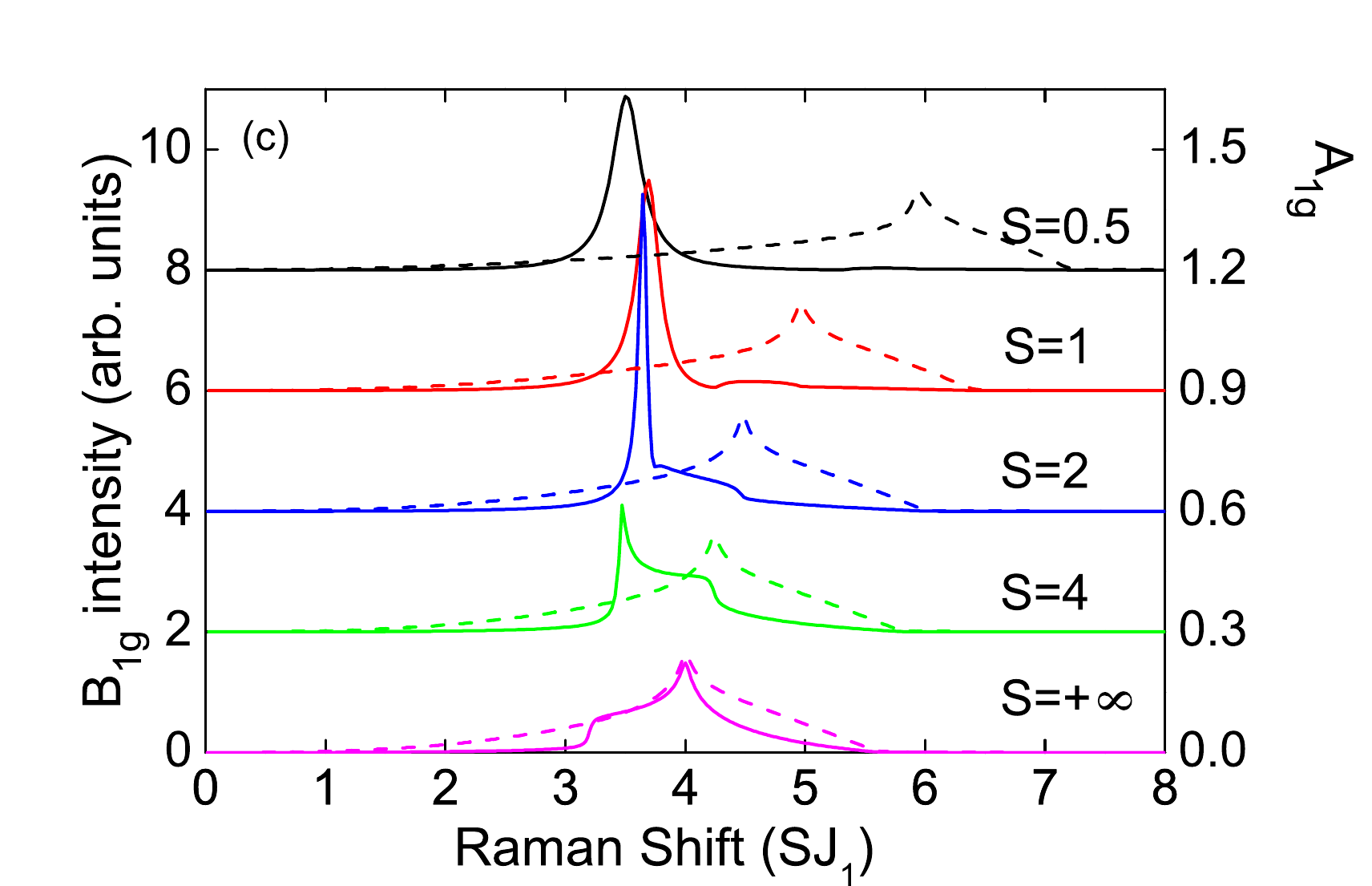}

\protect\caption{(Color online). Two-magnon Raman spectra for the $J_{1}-J_{2}$
model with $SJ_{1}=1$, $SJ_{2}=0.3$. (a)(b): spectra for $S=1$ highlighting the effects of Oguchi and m-m interaction terms.
(c): dependence of $S$ in $B_{1g}$ (solid lines) and $A_{1g}'$ (dashed lines) channels, both are calculated with the Oguchi and m-m interaction terms.}
\label{fig:NS}
\end{figure}

We consider the effects of quantum fluctuations beyond the LSW level. To focus on this point, we limit our discussion to the case $J_{1x}=J_{1y}=J_1$ and $J_c=0$ in this subsection. As is shown in Eq.~\eqref{eq:H1}, 
at the $1/S$ order, the corrections to the LSW results come from the following two parts:
the Oguchi term $H_0^\prime$ shifts the magnon dispersion to higher
energies, and the m-m interaction term $H_1^\prime$ allows repeat scattering of the light-excited magnon pairs. 
The roles of these two terms in two-magnon Raman spectra are 
shown in Fig.~\ref{fig:NS}.

The non-interacting (LSW) spectrum 
of the $B_{1g}$ channel typically shows a broad peak above an absorption edge at excitation energy $\omega\sim4J_1$(see Fig.~\ref{fig:NS}(a) and (b)).
When the m-m interaction is switched on, the spectral weight of this non-interacting
part is suppressed, and an additional peak below the absorption edge is developed.
This peak is particularly sharp when the spin size $S\lesssim4$, 
indicating
a resonance feature in this channel (Fig.~\ref{fig:NS}(c)). With increasing $S$, the position of the resonance peak is getting closer to the absorption edge, and its intensity is reduced, until eventually vanishes when $S\to\infty$. The existence of a sharp resonance peak makes the lineshape of the spectrum completely different once the m-m interaction is taken into account in the $B_{1g}$ channel. On the other hand, for the $A_{1g}^\prime$ channel, the spectrum is 
almost not modified by the m-m interaction. In fact, as we will discuss below, the peak in this channel is associated with a van Hove singularity in the density of states (DoS) of the one magnon dispersion, not affected by the m-m interaction (see Fig. \ref{fig:NJ2}(c)(d)). 

As shown in Fig.~\ref{fig:NS}(a) and (b), the Oguchi term slightly changes the lineshape of the spectrum. Its main effect is to 
push the spectral weight to higher energy. This explains well the monotonic shift (toward higher energy) of the peak position in the $A_{1g}^\prime$ channel as decreasing $S$ (Fig.~\ref{fig:NS}(c)). While, for $B_{1g}$ channel, the non-monotonic variance of the resonance peak position
as decreasing $S$ originates from the competition between the Oguchi term and m-m interactions.

\subsection{Role of $J_{2}$ frustration}

\begin{figure}
\includegraphics[bb=0bp 45bp 493bp 558bp,clip,scale=0.5]{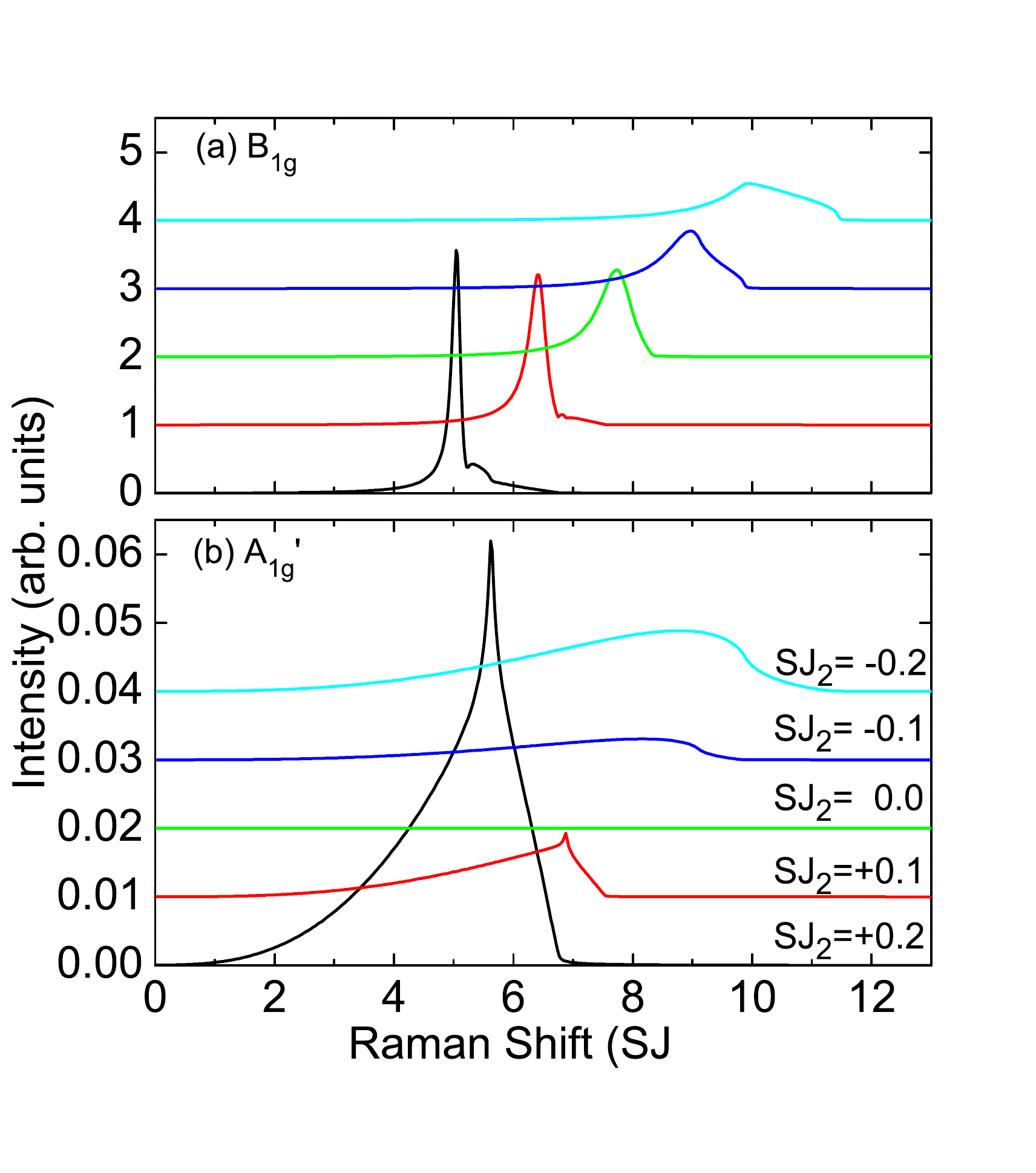}

\includegraphics[bb=0bp 0bp 493bp 449bp,clip,scale=0.5]{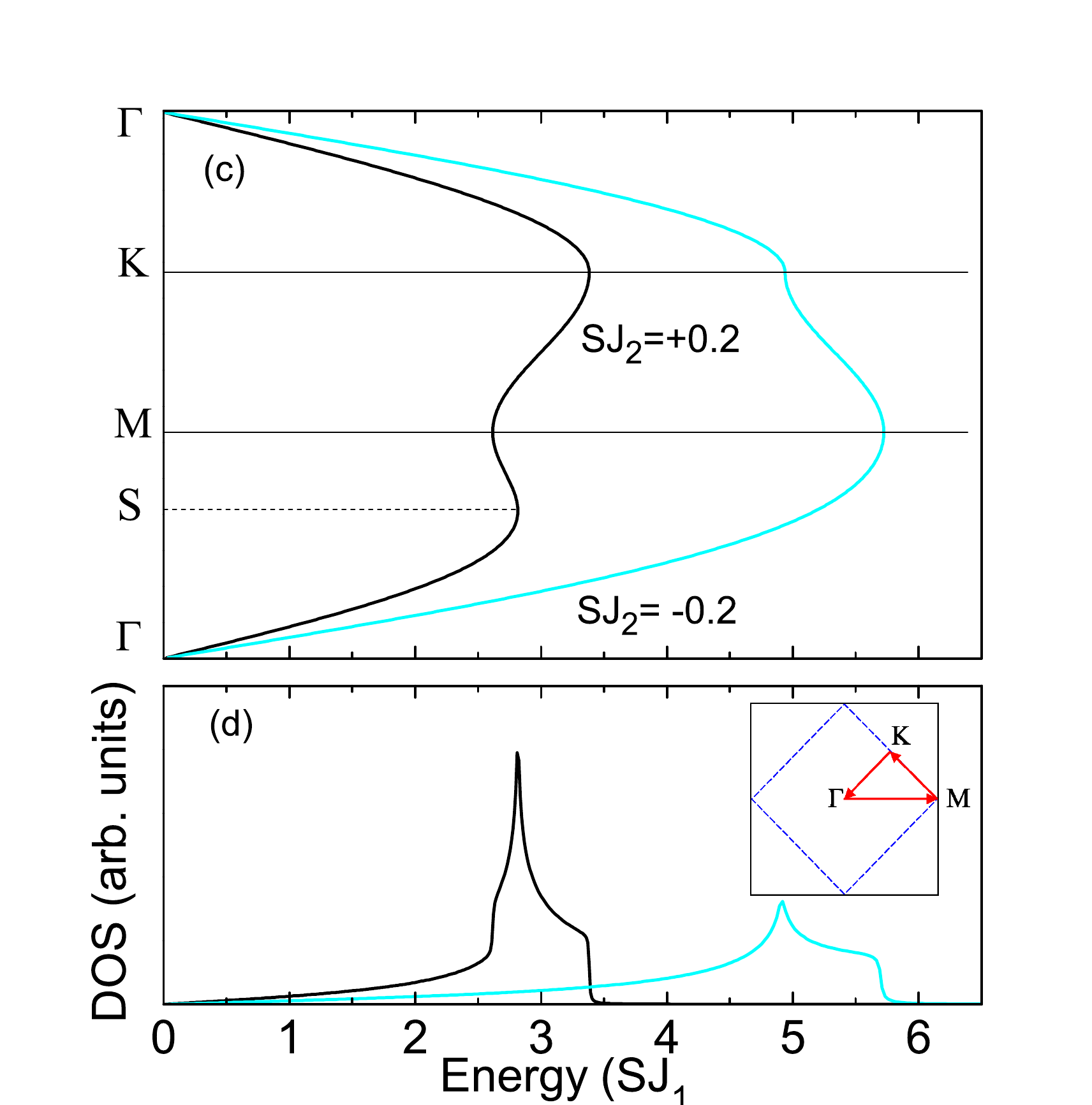}

\protect\caption{(Color online). (a)(b) Two-magnon Raman spectra for the frustrated $J_{1}-J_{2}$
model $SJ_{1}=1$, $S=2$ with different $J_{2}$; (c) magnon dispersion
and (d) one magnon DoS for $SJ_{1}=1$ $SJ_{2}=\pm0.2$, $S=2$.}
\label{fig:NJ2}
\end{figure}

The next nearest neighbor exchange coupling $J_2$ introduces frustration to the ground state. To study the effects of frustration, we consider the frustrated $J_{1}-J_{2}$
model with $SJ_{1}=1$ for various $J_{2}$ values. For $S=2$, the two-magnon
spectra in both $B_{1g}$ and $A_{1g}^\prime$ channels are shown in Fig. \ref{fig:NJ2}. We see that in the frustrated models with increasing antiferromagnetic $J_2$, the peaks of the Raman spectra in both channels become
sharper, where as in the non-frustrated models with ferromagnetic $J_2$, the spectral weights only show very broad humps. This difference can be understood by examining the one-magnon dispersion curves for the frustrated and non-frustrated models, shown in Fig.~\ref{fig:NJ2}(c), respectively. We 
find that the sharp resonance peak in the $B_{1g}$
channel in the frustrated case comes from resonant scattering 
of magnon pairs $\alpha_{\mathbf{k}}$
and $\beta_{\mathbf{-k}}$ near the $\mathbf{k}=(\pi,0)$ (M) point, where the dispersion has a local minimum. The resonant peak is completely suppressed when the dispersion turns to a local maximum at the M point in the non-frustrated model.

The peak in the $A_{1g}'$ channel, though evolves in a similar way as the resonance peak in the $B_{1g}$ channel, has a very different origin. 
In the frustrated case, we find there exists four saddle points in the dispersion along the $\Gamma$-M line. One of them is labeled as the S point in Fig.~\ref{fig:NJ2}(c). These saddle points contribute to a van Hove sigularity of the one-magnon DoS with a logarithmic divergence. 
This van Hove singularity contributes to the sharp $A_{1g}'$ peak. In the non-frustrated case, the saddle points along $\Gamma$-M line and the associated van Hove sigularity are removed, and hence the sharp peak does not appear in the $A_{1g}'$ channel. Note that besides the saddle points we just discussed, there can be two more (inequivalent) saddle points at $(\pi/2,\pi/2)$ and $(-\pi/2,\pi/2)$. But these saddle points do not contribute to singularities in the $A_{1g}'$ channel.

\subsection{Role of anisotropy}

\begin{figure}
\includegraphics[clip,scale=0.5]{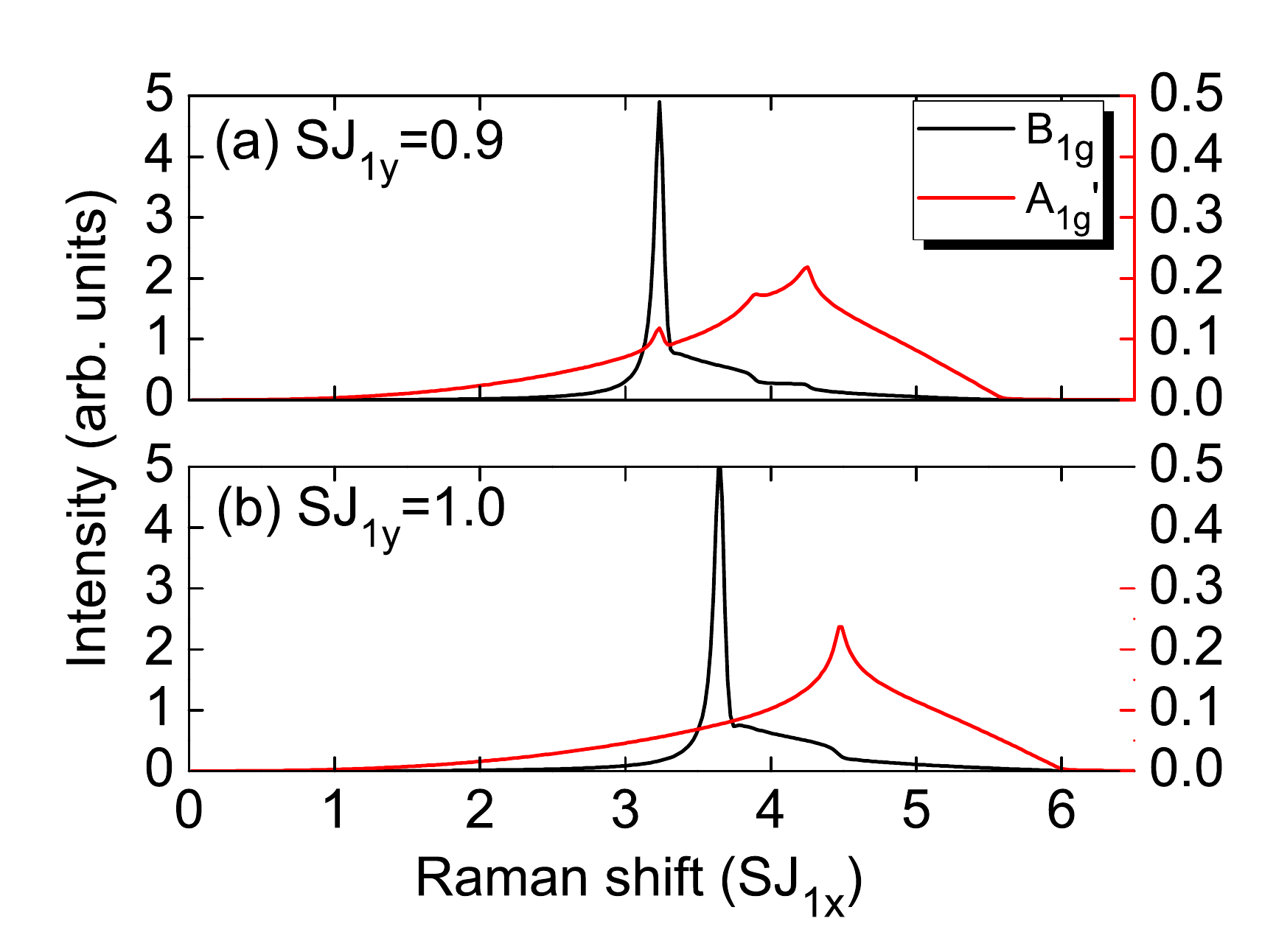}
\protect\caption{(Color online). Two-magnon Raman spectra for frustrated anisotropic
$J_{1x}-J_{1y}-J_{2}$ model with $SJ_{1x}=1$, $SJ_{1y}=0.9$, $SJ_{2}=0.3$,
$S=2$; (b) Reference isotropic model with same $S$, $J_{1x}$ and
$J_{2}$.}
\label{fig:NJ1y}
\end{figure}

We now study the effect of the exchange anisotropy, $J_{1x}\neq J_{1y}$, on the Raman scattering. This anisotropy serves as another perturbation to the $(\pi,\pi)$ AFM ground state. In the anisotropic $J_{1x}-J_{1y}-J_2$ model, the $D_{4h}$
symmetry is reduced to $D_{2h}$. As discussed above, the $B_{1g}$
and $A_{1g}'$ channels will share components with same irreducible
representations. 
In Fig.~\ref{fig:NJ1y} we show the results for $SJ_{1x}=1$, $SJ_{1y}=0.9$, $SJ_{2}=0.3$, and $S=2$. We see 
that the most significant changes of the spectrum by the anisotropy 
are in the $A_{1g}'$ channel: First, an additional peak in the $A_{1g}'$ channel emerges at the position of the resonance peak of the $B_{1g}$ channel. This 
behavior clearly indicates the two channels are not well separated when the $D_{4h}$ symmetry is broken. This peak is 
already visible when the anisotropy $J_{1y}/J_{1x}-1\gtrsim$ 5\%. Therefore, it
can be used to probe the possible exchange (and associated structural) anisotropy of materials as complementary to neutron diffraction. As another effect, the $A_{1g}'$ peak in the isotropic model is split which reflects the anisotropy of the one-magnon dispersion along the $(0,0)$-$(\pi,0)$ and $(0,0)$-$(0,\pi)$ directions.

\subsection{Role of the interlayer exchange coupling}

\begin{figure}
\includegraphics[clip,scale=0.5]{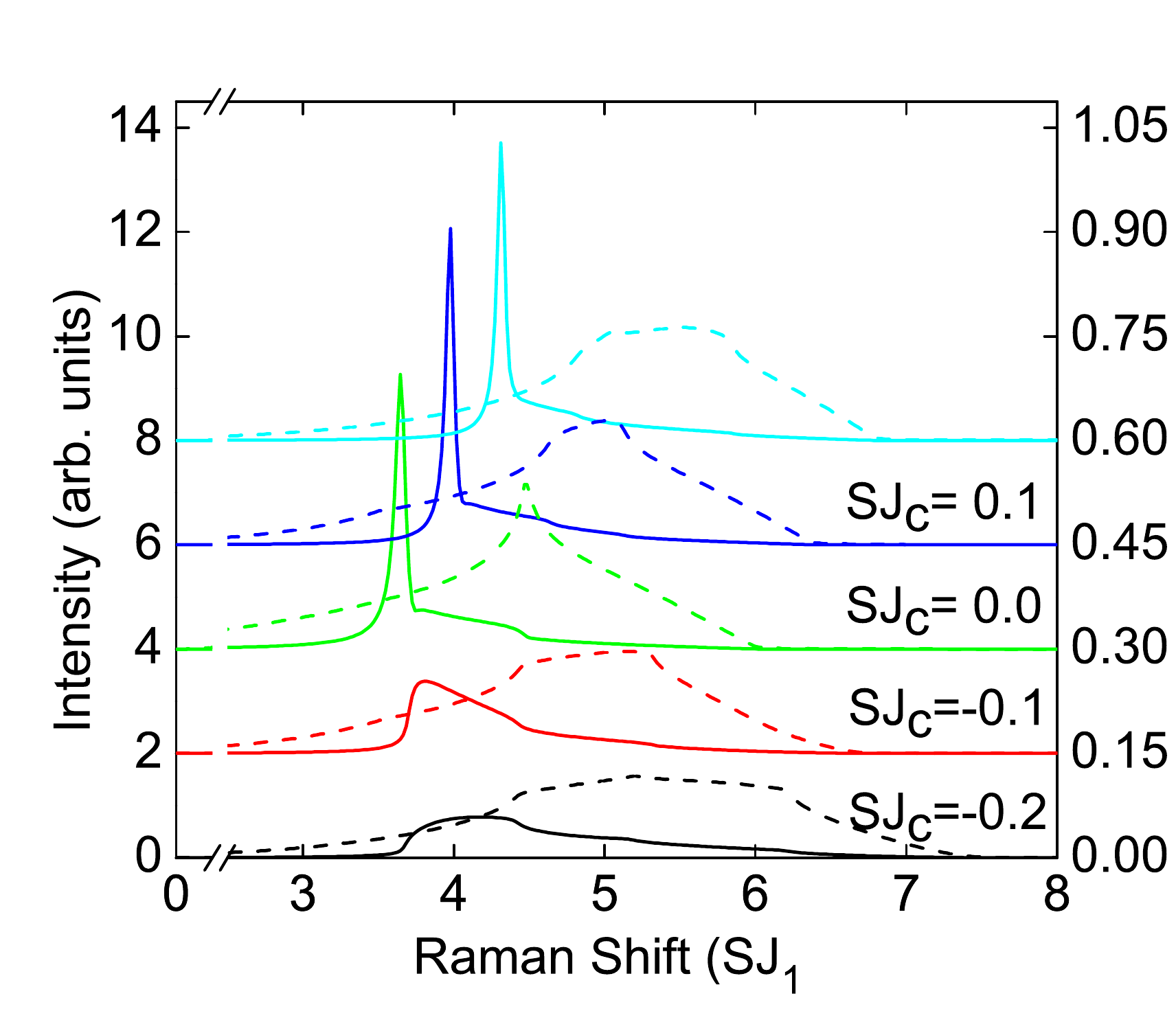}
\protect\caption{(Color online). $J_{c}$ dependence of two-magnon Raman spectra for
frustrated $J_{1}-J_{2}-J_{c}$ model $SJ_{1}=1$, $SJ_{2}=0.3$,
$S=2$. Solid lines represent $B_{1g}$ channel and dashed lines represent
$A_{1g}'$ channel.}
\label{fig:NJc}
\end{figure}

The previous results are discussed in 2D systems. In a more realistic 3D model,
interlayer exchange coupling $J_{c}$ will also 
affect the Raman spectra. Here we consider a $J_{1}-J_{2}-J_{c}$ model with $SJ_{1}=1$,
$SJ_{2}=0.3$, and $S=5/2$ for various $J_{c}$ values. The results are shown
in Fig. \ref{fig:NJc}. We see that for both FM and AFM $J_{c}$,
with increasing the magnitude of $J_c$, the sharp peak of the spectral weight in the $A_{1g}'$ channel evolves to a very broad platform. One can prove that at the LSW level, the width of this platform is proportional to the magnitude of $J_{c}$.

In the $B_{1g}$ channel it is remarkable that the sharp resonance peak
feature is suppressed significantly by a FM $J_{c}$ but preserves for an
AFM $J_{c}$. Such a phenomenon can be understood as follows: when
$J_{c}\neq0$, the magnons are dispersive along the $k_{z}$ direction. When $|J_{c}|$ is small, the magnon pair scattering
term $B_{1234}^{(3)}\alpha_{1}^{\dagger}\beta_{-4}^{\dagger}\beta_{-2}\alpha_{3}$
has little dependence on $k_{z}$, thus can be still treated as a 2D
process. The $k_{z}$ dependent magnon dispersion 
can be considered as an effective damping to the 2D system. 
Such an effective damping only affects the interacting part of the scattering cross section, and the bare part is not influenced.
The magnitude of this effective damping at the $(\pi,0)$ point is evaluated to be $\sim8SJ_{c}$ for the FM $J_{c}$ and $\sim SJ{}_{c}^{2}/(J_{1}-2J_{2})$ for the AFM $J_{c}$. We then see that the damping effect is much weaker for $J_{c}>0$ compared to the
$J_{c}<0$ systems. This explains why the resonance peak is robust for $J_c>0$, but are suppressed when $J_c<0$.

\section{Results For the $(\pi,0)$ Collinear order}

We also apply the same procedure to the case when the ground state has a $(\pi,0)$  collinear AFM order. Different
from the $(\pi,\pi)$ order, the collinear order intrinsically breaks the
$D_{4h}$ symmetry. From the discussion in the previous section,
the $B_{1g}$ and $A_{1g}'$ channels are not separated by symmetry and
share some common features in the spectra. Moreover, our calculation shows that the
intensity of the $A_{1g}'$ channel is about one order of magnitude higher than
the one in the $B_{1g}$ channel. So we will mainly focus on the spectrum of the $A_{1g}'$ channel in this section.

\begin{figure}
\includegraphics[bb=30bp 0bp 493bp 378bp,clip,scale=0.55]{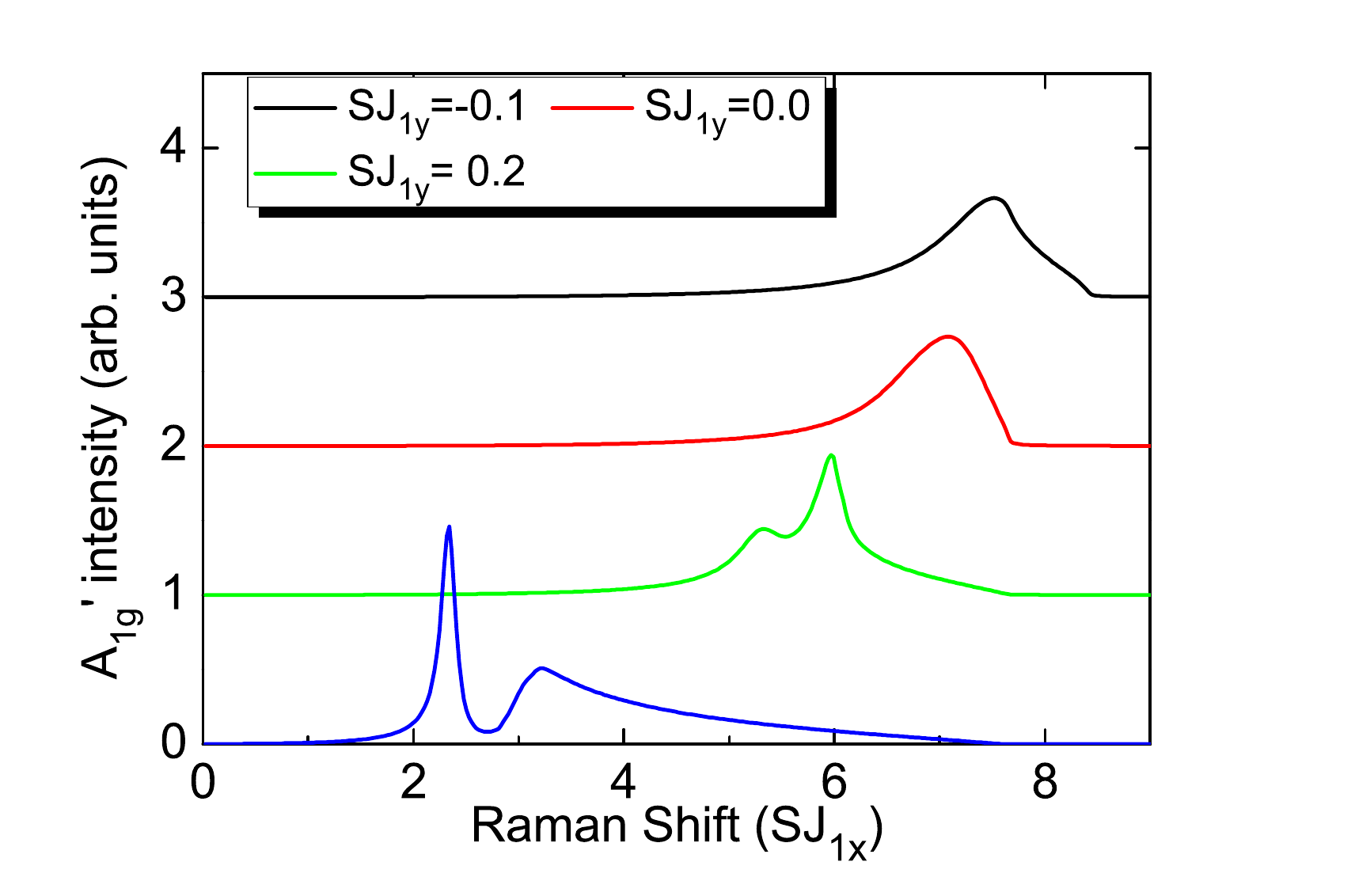}
\protect\caption{(Color online). $J_{1y}$ dependence of two-magnon Raman spectra for
anisotropic $J_{1x}-J_{1y}-J_{2}$ model with $SJ_{1x}=1$, $SJ_{2}=0.4$,
$S=1$.}
\label{fig:CJ1y}
\end{figure}

\begin{figure}
\includegraphics[bb=30bp 0bp 493bp 378bp,clip,scale=0.6]{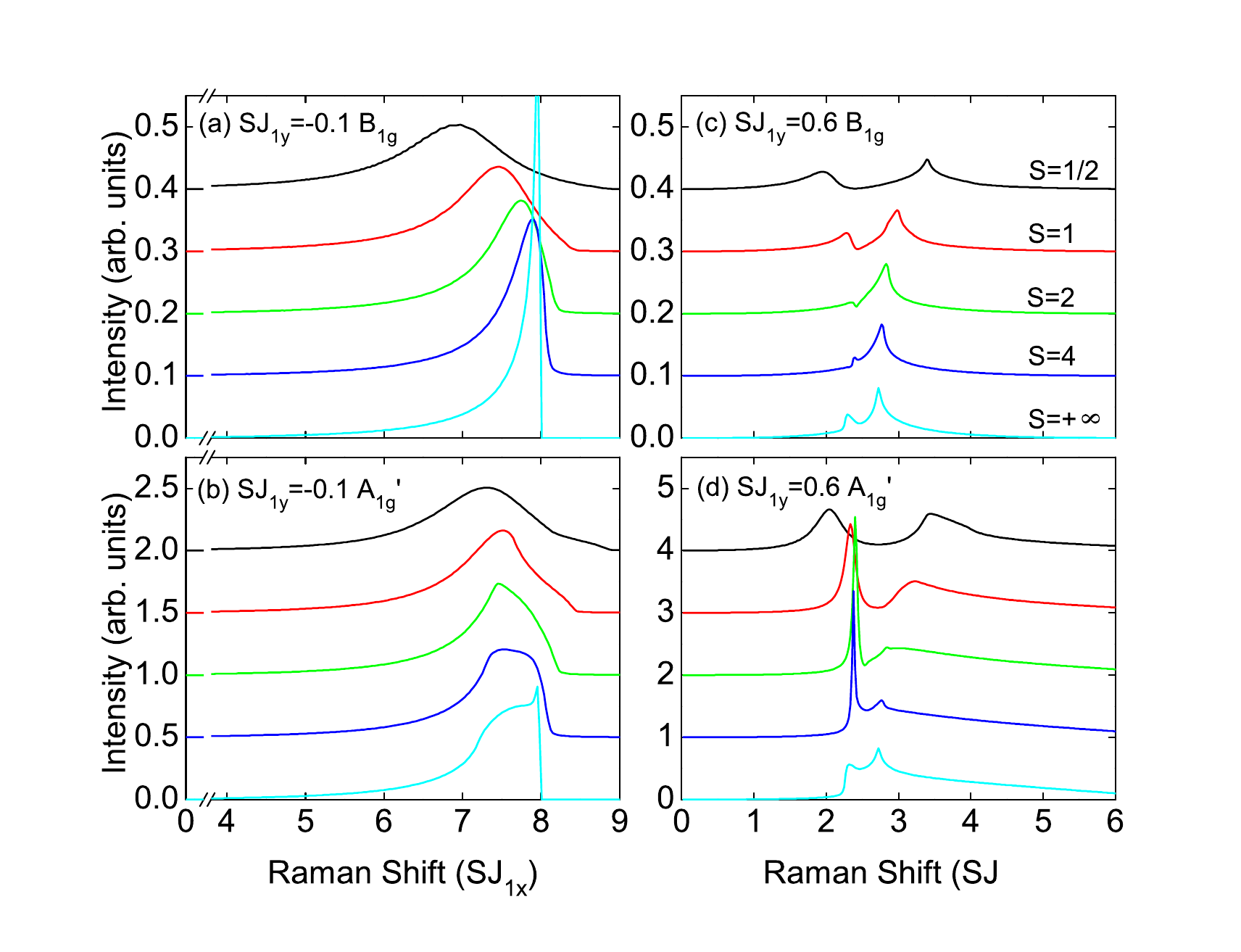}
\protect\caption{(Color online). $S$ dependence of two-magnon Raman spectra for anisotropic
$J_{1x}-J_{1y}-J_{2}$ model. (a),(b) unfrustrated model $SJ_{1x}=1$,
$SJ_{1y}=-0.1$, $SJ_{2}=0.4$; (c),(d) frustrated model $SJ_{1x}=1$,
$SJ_{1y}=0.6$, $SJ_{2}=0.4$.}
\label{fig:CS}
\end{figure}

The effects of frustration is shown in Fig. \ref{fig:CJ1y}. Similar
to the N\'{e}el ordered case, the frustration also pushes spectra to lower
energies. But the cut-off frequency is not strongly affected by the frustration.

In a non-frustrated model $J_{1y}<0$, the m-m interaction almost cancels
the peak in bare spectra completely and develops a peak at lower energy,
forming single broad peak structure in both channels. See \ref{fig:CS}
(a)(b). As the system becomes weakly frustrated, \textcolor{black}{in
$A_{1g}'$ channel the bare spectra at high energy will not be completely
canceled out by the m-m interaction, resulting a two-peak structure in this
channel.} The low energy peak, as is resulted by scattering resonance
only, is expected to vanish for large $S$. For highly frustrated
models, the two-peak structure emerges in both channels. We can see that
the two peaks survive even in the $S\rightarrow\infty$ limit, and they are pushed away from each other as $1/S$ increases, as
in shown in Fig. \ref{fig:CS} (c)(d). Note that in the $A_{1g}'$ channel,
the resonance peak becomes sharp when $S$ is about $1\sim4$. This
feature and its origin is very similar to the $B_{1g}$ channel in the N\'{e}el
ordered case.

Both positive and negative $J_{c}$ will shift the spectra to higher
frequencies. Also, as is already discussed in the previous section, the
interlayer coupling $J_{c}$ 
has the effect of a damping term to the 2D system. It is expected that sharp peak in corresponding
2D system can be damped by $J_{c}$.

\section{Discussions}

\subsection{Inplications for MnBi materials}

The material $\mathrm{BaMn_{2}Bi_{2}}$, which can be considered as the parent compound of the AMnBi$_2$ systems, is an AFM insulator with a large
ordered magnetic moment $\sim3.84\mu_{B}$ in each Mn ion. It has a similar structure to $A\mathrm{MnBi_{2}}$ except that the latter are metals consisting of a layer of Dirac electrons in the $A\mathrm{Bi}$ layer.

\begin{figure}
\includegraphics[clip,scale=0.5]{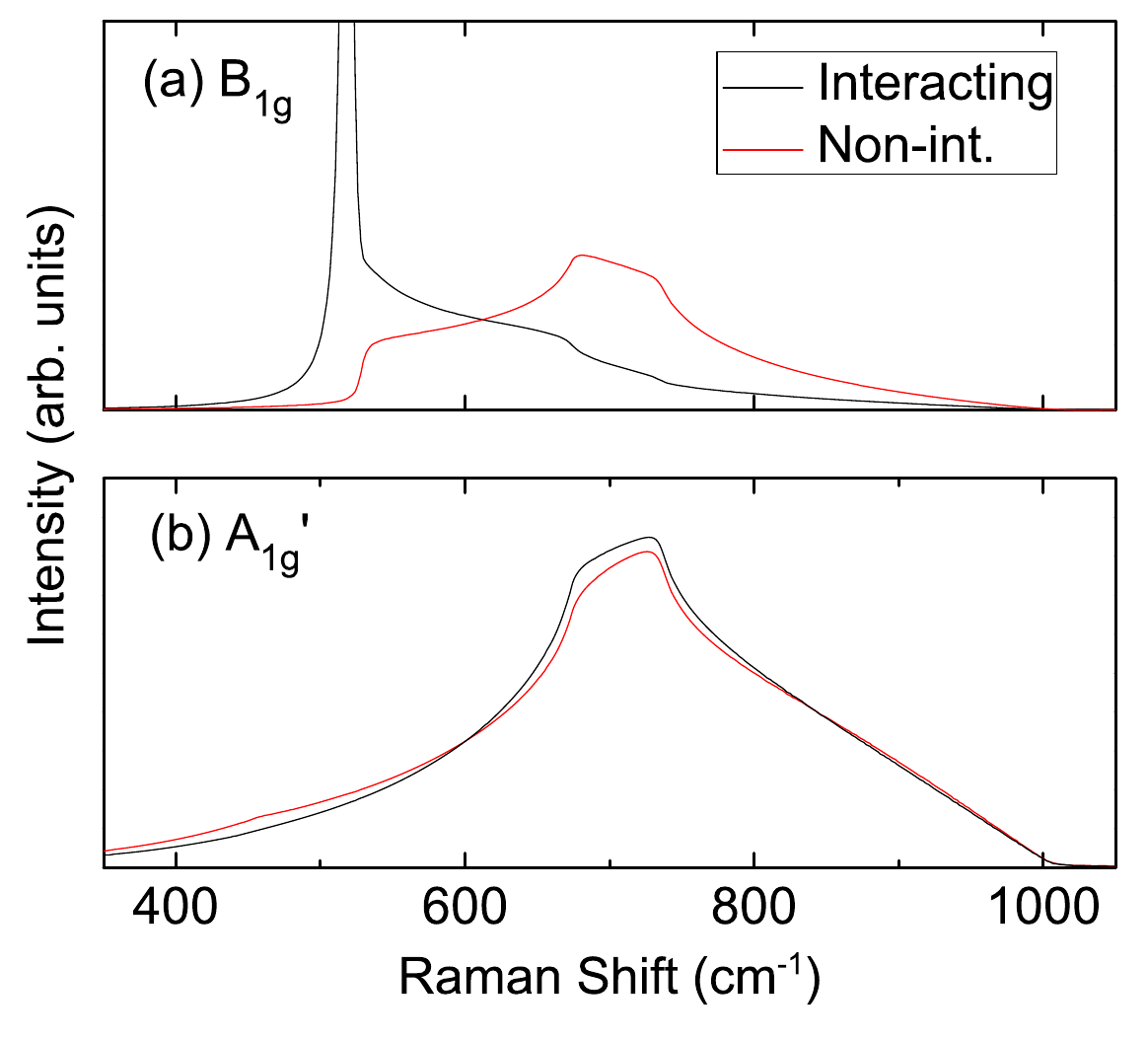}

\protect\caption{(Color online). Calculated two-magnon Raman spectra for $\mathrm{BaMn_{2}Bi_{2}}$ with $S=2$ and exchange parameters determined from inelastic neutron scattering experiment (see text).}

\label{fig:NThEx}
\end{figure}

We have calculated the two-magnon Raman spectra of a $J_{1}-J_{2}-J_{c}$ model for $\mathrm{BaMn_{2}Bi_{2}}$ using the exchange parameters $SJ_{1}=21.7(1.5)$, $SJ_{2}=7.85(1.4)$, $SJ_{c}=1.26(0.02)$, obtained from an inelastic neutron scattering experiment~\cite{calder2014magnetic}. We have taken the effective spin size to be $S=2$.
The result is presented in Fig. \ref{fig:NThEx}.
We see from the figure that a sharp resonance peak at wave number about 550 cm$^{-1}$ is present in the $B_{1g}$ channel. This indicates that for the model parameters taken, the effect of the m-m interaction can not be neglected, although the ordered moment is large.
We note that by measuring the peak positions in the $B_{1g}$ and $A_{1g}$ channels as well as the cut-off frequency, one may fully determine the exchange couplings of the system.
As for the Dirac materials
$A\mathrm{MnBi_{2}}$, We expect similar Raman spectra, given that the spin dynamics is dominant by the interacting local moments. The itinerant electrons may contribute additional damping to the resonance peak in the $B_{1g}$ channel, and may also renormalize the values of the exchange couplings via the induced RKKY interactions.

\subsection{Iron based materials}

\begin{figure}
\includegraphics[clip,scale=0.3]{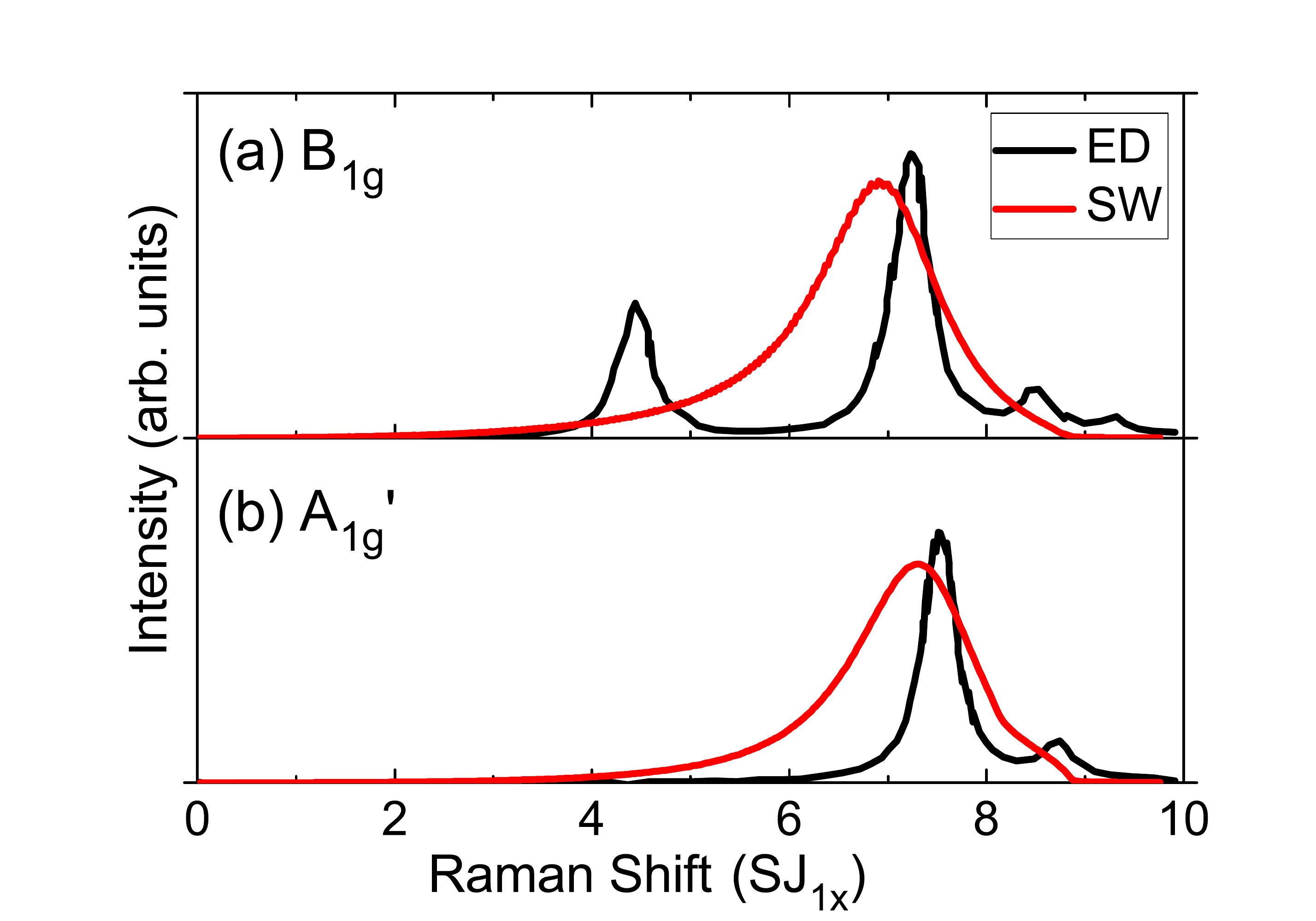}
\protect\caption{(Color online). Comparison of ED and spin wave approaches to two-magnon
Raman spectra in anisotropic $J_{1x}-J_{1y}-J_{2}$ model with $SJ_{1x}=1$,
$SJ_{1y}=-0.1$, $SJ_{2}=0.4$, $S=1/2$. (a) $B_{1g}$ channel; (b)
$A_{1g}'$ channel. }
\label{fig:CED}
\end{figure}

As there is $S=1/2$ ED calculation result for $\mathrm{CaFe_{2}As_{2}}$($J_{1y}=-0.1J_{1x}$,
$J_{2}=0.4J_{1x}$ as reported by INS measurements) available\cite{chen2011theory},
we make a comparison with our spectra and their 36 sites ED result,
as shown in Fig. \ref{fig:CED}. Our peak position is in consistent
with theirs, but the details of spectral lineshape is entirely different.
What we obtain is a single broad peak structure for both polarizations.
Our lineshape should be more reliable than the ED result for the presence
of strong finite-size effects in the ED calculation.

Since for each Fe ion, there are two degenerate $3d$ orbitals, namely $d_{xz}$
and $d_{yz}$ orbitals, to be active, we can also consider an $S=1$ spectrum
for the same $SJ$. The spectra are shown in Fig. \ref{fig:CS}
(a)(b) (red lines). We find that one broad peak structrure in both
channels remains, and the peaks are slightly shifted to higher frequencies
compared with the $S=1/2$ case. The interlayer coupling $J_{c}$ further
shifts the peak to higher frequencies. Taking $SJ_{1x}\sim50$meV
and $SJ_{1c}\sim5$meV as is reported in INS experiments, we expect
a broad two-magnon peak at $7.9SJ_{1x}\sim3185\mathrm{cm^{-1}}$.

\section{Concluding Remarks}

In this paper we have made a comprehensive study of two-magnon Raman
spectra in $J_{1x}-J_{1y}-J_{2}-J_{c}$ antiferromagnets using spin
wave theory. Our treatment includes the contribution to m-m interactions
at the $1/S$ order. The m-m interactions are taken into account within
the ladder approximation, and the ladder diagrams are summed up exactly.

We find that for isotropic Neel ordered system, $B_{1g}$ and $A_{1g}'$
channels are well separated by symmetry. $B_{1g}$ channel is strongly
modified by m-m interaction while $A_{1g}'$ channel is not. We predict
that for large $S$ system, in $B_{1g}$ channel, sharp resonance
peak emerges in frustrated systems when $J_{c}\geq0$ and the peak
is suppressed for ferromagnetic $J_{c}$. In $A_{1g}'$ channel, a
platform can be opened by $J_{c}$. For anisotropic system the $B_{1g}$
resonance peak will tunnel into $A_{1g}'$ channel. For collinear
ordered system, we predict one peak structrure for non-frustrated
systems, and the peak splits due to frustration.

Our results suggest that the two-magnon Raman spectra can be used to probe the exchange anisotropy, which serves as complementary to inelastic neutron scattering.
With Raman, $J$ values can be determined by characteristic
frequencies which correspond to van-Hove singularities in one-magnon
DoS, since these frequencies are not shifted by m-m interactions at
the $1/S$ order. Anisotropy in Néel ordered frustrated system is manifested
in tunneling of the $B_{1g}$ resonance peak into $A_{1g}'$ channel.
$J$'s obtained by Raman is expected to be more accurate than INS's
due to its higher resolution. 

\appendix

\section{Parameters}

The definition of FBZ are shown in Table \ref{tab:FBZ}. Note that
our FBZ has spatial inversion symmetry, which is essential for simplifying
the Oguchi's term $A_{\mathbf{k}}$.

\begin{table}
\protect\caption{\label{tab:FBZ} FBZ}

\begin{ruledtabular}
\begin{tabular}{ccc}
 & Néel order & Collinear order\tabularnewline
\colrule$J_{c}\leq0$ & $|ak_{x}|+|ak_{y}|\leq\pi\,\,\,|ck_{z}|\leq\pi$ & $|2ak_{x}|,|bk_{y}|,|ck_{z}|\leq\pi$\tabularnewline
\multirow{2}{*}{$J_{c}>0$} & $|ak_{x}|,|ak_{y}|,|ck_{z}|\leq\pi$ & $|bk_{y}|\leq\pi$\tabularnewline
 & $|ak_{x}|+|ak_{y}|+|ck_{z}|\leq\frac{3}{2}\pi$ & $|ak_{x}|+|\frac{b}{2}k_{y}|+|ck_{z}|\leq\frac{3}{2}\pi$\tabularnewline
\end{tabular}\end{ruledtabular}

\end{table}

The quantity $P_{\mathbf{k}}$, $Q_{\mathbf{k}}$, $A_{\mathbf{k}}$
and $B_{1234}^{(3)}$ can be written as sum of contributions from
each bond: $P_{\mathbf{k}}=\sum_{b}J_{b}z_{b}P_{b,\mathbf{k}}$, $Q_{\mathbf{k}}=\sum_{b}J_{b}z_{b}Q_{b,\mathbf{k}}$,
$A_{\mathbf{k}}=\sum_{b}J_{b}z_{b}A_{b,\mathbf{k}}$, $B_{1234}^{(3)}=\sum_{b}J_{b}z_{b}B_{b,1234}^{(3)}$,
where $b$ is the type of bonds, which runs over the set $\{1x,1y,2,c\}$,
and $z_{b}$ is the coordination number of bond $b$. The definition
of $P_{b,\mathbf{k}}$, $Q_{b,\mathbf{k}}$, and $A_{b,\mathbf{k}}$
are shown in Table \ref{tab:PQ}. $B_{b,1234}^{(3)}=-\{\gamma_{b\:2-4}+\gamma_{b\:1-3}x_{1}x_{2}x_{3}x_{4}+\gamma_{b\:1-4}x_{1}x_{2}+\gamma_{b\:2-3}x_{3}x_{4}-\frac{1}{2}[\gamma_{b\,2}x_{4}+\gamma_{b\,1}x_{1}x_{2}x_{4}+\gamma_{b\,2-3-4}x_{3}+\gamma_{b\,1-3-4}x_{1}x_{2}x_{3}+\gamma_{b\,4}x_{2}+\gamma_{b\,3}x_{2}x_{3}x_{4}+\gamma_{b\,4-2-1}x_{1}+\gamma_{b\,3-2-1}x_{1}x_{3}x_{4}]\}$
for AFM bond $b$ and $B_{b,1234}^{(3)}=\frac{1}{2}(\gamma_{b\,1-3}+\gamma_{b\,1-4}+\gamma_{b\,2-3}+\gamma_{b\,2-4}-\gamma_{b\,1}-\gamma_{b\,2}-\gamma_{b\,3}-\gamma_{b\,4})(x_{2}x_{4}+\mathrm{sgn}\gamma_{\mathbf{G}}x_{1}x_{3})$
for FM bond $b$. Here use the notation $\gamma_{1x\mathbf{k}}=\cos k_{x}a$,
$\gamma_{1y\mathbf{k}}=\cos k_{y}a$, $\gamma_{2\mathbf{k}}=\cos k_{x}a\cos k_{y}a$,
$\gamma_{c\mathbf{k}}=\cos k_{z}a$.

It should be noticed that here we define a bond ferromagnetic, when
the bond is connecting sites in the same sublattice, or antiferromagnetic
otherwise. (It does not directly depend on the sign of exchange parameter
of the bond.)

\begin{table}
\protect\caption{\label{tab:PQ} Definition of $P_{b,\mathbf{k}}$, $Q_{b,\mathbf{k}}$
and $A_{b,\mathbf{k}}$}

\begin{ruledtabular}
\begin{tabular}{ccc}
 & AFM bond & FM bond\tabularnewline
\colrule$P_{b,\mathbf{k}}$ & $S$ & $-S(1-\gamma_{b\mathbf{k}})$\tabularnewline
$Q_{b,\mathbf{k}}$ & $S\gamma_{b\mathbf{k}}$ & $0$\tabularnewline
$A_{b,\mathbf{k}}$%
\footnote{The original form of $A_{b,\mathbf{k}}$ can be written as $A_{b,\mathbf{k}}=\frac{2}{N}\sum_{\mathbf{p}}\frac{1}{2\epsilon_{\mathbf{k}}\epsilon_{\mathbf{p}}}[-1+\epsilon_{\mathbf{p}}+\gamma_{\mathbf{p}}\gamma_{b\mathbf{p}}+\gamma_{b\mathbf{k}}(\gamma_{\mathbf{k}}-\epsilon_{\mathbf{p}}\gamma_{\mathbf{k}}-\gamma_{\mathbf{p}}\gamma_{b\,\mathbf{p-k}})]$
for AFM bond and $A_{b,\mathbf{k}}=\frac{2}{N}\sum_{\mathbf{p}}\frac{1-\epsilon_{\mathbf{p}}}{2\epsilon_{\mathbf{k}}\epsilon_{\mathbf{p}}}(\gamma_{b\,\mathbf{k-p}}-\gamma_{b\mathbf{k}}-\gamma_{b\mathbf{p}}+1)$
for FM bond. It is simplified by using the equality $\sum_{\mathbf{p}}\gamma_{b\,\mathbf{p-k}}=\sum_{\mathbf{p}}\gamma_{b\mathbf{k}}\gamma_{b\mathbf{p}}$
which holds when FBZ has spatial inversion symmetry.%
} & $\frac{1}{2\epsilon_{k}}(1-\gamma_{\mathbf{k}}\gamma_{b\mathbf{k}})A_{b}$ & $\frac{1}{2\epsilon_{k}}(1-\gamma_{b\mathbf{k}})A_{b}$\tabularnewline
$A_{b}$ & $\frac{2}{N}\sum_{\mathbf{p}}\frac{\gamma_{\mathbf{p}}\gamma_{b\mathbf{p}}+\epsilon_{\mathbf{p}}-1}{\epsilon_{\mathbf{p}}}$ & $\frac{2}{N}\sum_{\mathbf{p}}\frac{1-\epsilon_{\mathbf{p}}-\gamma_{b\mathbf{p}}}{\epsilon_{\mathbf{p}}}$\tabularnewline
\end{tabular}\end{ruledtabular}

\end{table}

Values of $M_{\mathbf{k}}$ are shown in table \ref{tab:Mk}.

\begin{table*}[b]
\protect\caption{\label{tab:Mk} Definition of $M_{\mathbf{k}}$}

\begin{ruledtabular}
\begin{tabular}{ccc}
Polarization & Néel order & Collinear order\tabularnewline
\colrule$B_{1g}$ & $\sqrt{\frac{2}{N}}\frac{S}{2\epsilon_{k}}[J_{1x}z_{1x}(\gamma_{1x\mathbf{k}}-\gamma_{\mathbf{k}})-J_{1y}z_{1y}(\gamma_{1y\mathbf{k}}-\gamma_{\mathbf{k}})]$ & $\sqrt{\frac{2}{N}}\frac{S}{2\epsilon_{k}}[J_{1x}z_{1x}(\gamma_{1x\mathbf{k}}-\gamma_{\mathbf{k}})-J_{1y}z_{1y}\gamma_{\mathbf{k}}(1-\gamma_{1y\mathbf{k}})]$\tabularnewline
\multirow{2}{*}{$A_{1g}^{'}$} & $\sqrt{\frac{2}{N}}\frac{S}{2\epsilon_{k}}\{J_{1x}z_{1x}(\gamma_{1x\mathbf{k}}-\gamma_{\mathbf{k}})+J_{1y}z_{1y}(\gamma_{1y\mathbf{k}}-\gamma_{\mathbf{k}})$ & $\sqrt{\frac{2}{N}}\frac{S}{2\epsilon_{k}}\{J_{1x}z_{1x}(\gamma_{1x\mathbf{k}}-\gamma_{\mathbf{k}})+J_{1y}z_{1y}\gamma_{\mathbf{k}}(1-\gamma_{1y\mathbf{k}})$\tabularnewline
 & $+4J_{2}\times2\times\gamma_{\mathbf{k}}[1-\cos(k_{x}a+k_{y}a)]\}$ & $+4J_{2}\times2\times[\cos(k_{x}a+k_{y}a)-\gamma_{\mathbf{k}}]\}$\tabularnewline
\end{tabular}\end{ruledtabular}

\end{table*}

The channels $v_{n}(\mathbf{k})$ in Néel and Collinear ordered phase
are defined in Table \ref{tab:vnk}.

\begin{table}[b]
\protect\caption{\label{tab:vnk} Definition of the channels $v_{n\mathbf{k}}$}

\begin{ruledtabular}
\begin{tabular}{ccc}
n & Néel & Collinear\tabularnewline
\colrule1 & $l_{\mathbf{k}}^{2}\cos k_{x}$ & $l_{\mathbf{k}}^{2}\cos k_{x}$\tabularnewline
2 & $l_{\mathbf{k}}^{2}\sin k_{x}$ & $l_{\mathbf{k}}^{2}\sin k_{x}$\tabularnewline
3 & $m_{\mathbf{k}}^{2}\cos k_{x}$ & $m_{\mathbf{k}}^{2}\cos k_{x}$\tabularnewline
4 & $m_{\mathbf{k}}^{2}\sin k_{x}$ & $m_{\mathbf{k}}^{2}\sin k_{x}$\tabularnewline
5 & $l_{\mathbf{k}}^{2}\cos k_{y}$ & $l_{\mathbf{k}}^{2}\cos k_{x}\cos k_{y}$\tabularnewline
6 & $l_{\mathbf{k}}^{2}\sin k_{y}$ & $l_{\mathbf{k}}^{2}\sin k_{x}\cos k_{y}$\tabularnewline
7 & $m_{\mathbf{k}}^{2}\cos k_{y}$ & $l_{\mathbf{k}}^{2}\cos k_{x}\sin k_{y}$\tabularnewline
8 & $m_{\mathbf{k}}^{2}\sin k_{y}$ & $l_{\mathbf{k}}^{2}\sin k_{x}\sin k_{y}$\tabularnewline
9 & $l_{\mathbf{k}}m_{\mathbf{k}}\cos k_{x}\cos k_{y}$ & $m_{\mathbf{k}}^{2}\cos k_{x}\cos k_{y}$\tabularnewline
10 & $l_{\mathbf{k}}m_{\mathbf{k}}\sin k_{x}\cos k_{y}$ & $m_{\mathbf{k}}^{2}\sin k_{x}\cos k_{y}$\tabularnewline
11 & $l_{\mathbf{k}}m_{\mathbf{k}}\cos k_{x}\sin k_{y}$ & $m_{\mathbf{k}}^{2}\cos k_{x}\sin k_{y}$\tabularnewline
12 & $l_{\mathbf{k}}m_{\mathbf{k}}\sin k_{x}\sin k_{y}$ & $m_{\mathbf{k}}^{2}\sin k_{x}\sin k_{y}$\tabularnewline
13 & $l_{\mathbf{k}}m_{\mathbf{k}}$ & $l_{\mathbf{k}}m_{\mathbf{k}}\cos k_{y}$\tabularnewline
14 & $l_{\mathbf{k}}m_{\mathbf{k}}\cos k_{z}$ & $l_{\mathbf{k}}m_{\mathbf{k}}\sin k_{y}$\tabularnewline
15 & $l_{\mathbf{k}}m_{\mathbf{k}}\sin k_{z}$ & $l_{\mathbf{k}}m_{\mathbf{k}}$\tabularnewline
16 & $m_{\mathbf{k}}^{2}\cos k_{z}$ & $l_{\mathbf{k}}m_{\mathbf{k}}\cos k_{z}$\tabularnewline
17 & $m_{\mathbf{k}}^{2}\sin k_{z}$ & $l_{\mathbf{k}}m_{\mathbf{k}}\sin k_{z}$\tabularnewline
18 & $l_{\mathbf{k}}^{2}\cos k_{z}$ & $m_{\mathbf{k}}^{2}\cos k_{z}$\tabularnewline
19 & $l_{\mathbf{k}}^{2}\sin k_{z}$ & $m_{\mathbf{k}}^{2}\sin k_{z}$\tabularnewline
20 & - & $l_{\mathbf{k}}^{2}\cos k_{z}$\tabularnewline
21 & - & $l_{\mathbf{k}}^{2}\sin k_{z}$\tabularnewline
\end{tabular}\end{ruledtabular}

\end{table}

Matrix elements of $\hat{\Gamma}$ in Néel order is given by

$\hat{\Gamma}=\frac{2}{N}\left(\begin{array}{ccc}
\hat{X} & \hat{U}\\
\hat{U}^{T} & W & \hat{V}\\
 & \hat{V}^{T} & \hat{Z}
\end{array}\right)$

where definition of submatrices are shown in Table \ref{tab:gamma}.
\begin{widetext}
\begin{table*}[b]
\protect\caption{\label{tab:gamma} Definition of submatrices in $\hat{\Gamma}$}

\begin{ruledtabular}
\begin{tabular}{ccc}
 & Néel & Collinear\tabularnewline
\colrule$\hat{X}$ & $-2\begin{pmatrix}J_{1x}\hat{\mathbf{1}}_{4\times4}\\
 & J_{1y}\hat{\mathbf{1}}_{4\times4}\\
 &  & -4J_{2}\hat{\mathbf{1}}_{4\times4}
\end{pmatrix}$ & $-2\begin{pmatrix}J_{1x}\hat{\mathbf{1}}_{4\times4}\\
 & 2J_{2}\hat{\mathbf{1}}_{8\times8}\\
 &  & -2J_{1y}\hat{\mathbf{1}}_{2\times2}
\end{pmatrix}$\tabularnewline
$\hat{U}$ & $-2\left(\begin{array}{cccccccccccc}
J_{1x} & 0 & J_{1x} & 0 & J_{1y} & 0 & J_{1y} & 0 & 4J_{2} & 0 & 0 & 0\end{array}\right)^{T}$ & $-2\left(\begin{array}{cccccccccccccc}
J_{1x} & 0 & J_{1x} & 0 & 2J_{2} & 0 & 0 & 0 & 2J_{2} & 0 & 0 & 0 & 2J_{1y} & 0\end{array}\right)^{T}$\tabularnewline
$W$ & $-4J_{1x}-4J_{1y}+8J_{2}-4|J_{c}|$ & $-4J_{1x}+4J_{1y}-8J_{2}-4|J_{c}|$\tabularnewline
\multirow{2}{*}{$\hat{V}$} & \multicolumn{2}{c}{$-4J_{c}\left(\begin{array}{cccccc}
1 & 0 & 0 & 0 & 0 & 0\end{array}\right)$ $J_{c}\leq0$}\tabularnewline
 & \multicolumn{2}{c}{$-2J_{c}\left(\begin{array}{cccccc}
0 & 0 & 1 & 0 & 1 & 0\end{array}\right)$ $J_{c}>0$}\tabularnewline
\multirow{2}{*}{$\hat{Z}$} & \multicolumn{2}{c}{$4J_{c}\left(\begin{array}{cc}
\hat{\mathbf{1}}_{2\times2}\\
 & \hat{\mathbf{0}}_{4\times4}
\end{array}\right)$ $J_{c}\leq0$}\tabularnewline
 & \multicolumn{2}{c}{$-2J_{c}\left(\begin{array}{cc}
\hat{\mathbf{0}}_{2\times2}\\
 & \hat{\mathbf{1}}_{4\times4}
\end{array}\right)$ $J_{c}>0$}\tabularnewline
\end{tabular}\end{ruledtabular}

\end{table*}

\end{widetext}

\bibliographystyle{apsrev4-1}
\nocite{*}

%
\end{document}